\documentclass{elsart}
\usepackage{epsfig}
\usepackage{multirow}
\newcounter{subfigure}

\begin{document}
\begin{frontmatter}

\title{Nuclear structure of $^{231}$Ac}

\author[Madrid]{R.~Boutami},
\author[Madrid]{M.J.G.~Borge\corauthref{cor}},
\author[Uppsala]{H.~Mach},
\author[Warsaw]{W.~Kurcewicz},
\author[Madrid2,Isolde]{L.M.~Fraile},
\author[Warsaw]{K.~Gulda},
\author[Oslo1]{A.J.~Aas},
\author[Valencia]{L.M.~Garc\'{\i}a-Raffi},
\author[Oslo2]{G.~L{\o}vh{\o}iden},
\author[Valencia]{T.~Mart\'{\i}nez},
\author[Valencia]{B.~Rubio},
\author[Valencia]{J.L.~Ta\'{\i}n} and
\author[Madrid,Isolde]{O.~Tengblad}

\corauth[cor]{borge@iem.cfmac.csic.es}

\address[Madrid]{Instituto de Estructura de la Materia, CSIC,
Serrano 113 bis, E-28006 Madrid, Spain}
\address[Uppsala]{Department of Radiation Sciences,
ISV, Uppsala University, SE-751 21 Uppsala, Sweden}
\address[Warsaw]{Department of Physics, University of Warsaw, Pl-00 681 Warsaw,
Poland}
\address[Madrid2]{Departamento F\'{\i}sica At\'omica, Molecular y Nuclear, Facultad CC. F\'{\i}sicas, Universidad Complutense, E-28040 Madrid, Spain}
\address[Isolde]{ISOLDE, PH Department, CERN, CH-1211 Geneva 23, Switzerland}
\address[Oslo1]{Department of Chemistry, University of Oslo, P.O.Box 1033, Blindern,
N-0315 Oslo, Norway}
\address[Valencia]{Instituto de F\'{\i}sica Corpuscular, CSIC --
Universidad de Valencia, Apdo. 22805, E-46071 Valencia, Spain}
\address[Oslo2]{Department of Physics, University of Oslo, P.O.Box 1048, Blindern, N-0316 Oslo, Norway}

\begin{abstract}
The low-energy structure of $^{231}$Ac has been investigated by
means of $\gamma$ ray spectroscopy following the $\beta^-$ decay of
$^{231}$Ra. Multipolarities of 28 transitions have been established
by measuring conversion electrons with a \textsc{mini-orange} electron
spectrometer. The decay scheme of $^{231}$Ra $\rightarrow$
$^{231}$Ac has been constructed for the first time. The Advanced Time
Delayed $\beta\gamma\gamma(t)$ method has been used to measure the
half-lives of five levels.
The moderately fast B(E1) transition rates derived suggest that the octupole effects,
albeit weak, are still present in this exotic nucleus.



\end{abstract}
\end{frontmatter}

\section{Introduction}
Since the observation of low-lying K$^{\pi }$=0$^-$ bands in doubly even radium
nuclei~\cite{ste54} the possibility that some nuclei can be described
by a mean field approach with broken reflection symmetry has been considered~\cite{ald56}.
Numerous experimental and theoretical discoveries
were done in the 80's providing extra evidence of reflection
asymmetric octupole deformation around A=225~\cite{ahm93}.
Many of these nuclei have been studied at ISOLDE in beta decay experiments~\cite{kur00}.
In odd-A nuclei an important feature of static octupole deformation
is the observation
of parity doublets, i.e. rotational bands lying close in excitation energy
with the same intrinsic parameters and spins, but opposite parity.

A survey of the available data for odd-A nuclei reveals that the largest
amount of octupole correlations is present in the Ac and Pa nuclei.
In particular for the odd actinium isotopes near A = 225
spectroscopy studies have been done in $\alpha $ decay ($^{219-225}$Ac), $\beta$ decay ($^{225-231}$Ac)
and reactions (see~\cite{fir99} and references therein).
The high spin states of $^{219}$Ac observed in heavy-ion fusion reactions~\cite{dri86} have been
described using small quadrupole and octupole deformation components
$\beta_2$ = $\beta_3$ = 0.05.
The nucleus $^{221}$Ac was predicted as a good example of octupole
instability with the K = 5/2 and K= 3/2 band heads calculated to be very close in energy.
Experimentally two bands were found~\cite{aic94} with the same angular momentum,
opposite parity and an average difference of 44 keV between equivalent members of the doublets.
This fact suggested parity doublets but the differences in intrinsic dipole moment (D$_0$)
values indicated that the two bands originate from different orbitals.
The low-energy spectra of $^{223,225}$Ac display the  K=3/2 and K=5/2
parity doublets~\cite{ahm93,cwi91}. In $^{227}$Ac only the K=5/2
parity doublet band, observed at 300 keV
excitation energy, can be described in terms of static octupole deformation.
In the case of $^{229}$Ac the four K=1/2$^{\pm} $ and K=3/2$^{\pm} $  bands
observed in (t,$\alpha $) reactions~\cite{tho77}
are interpreted either as normal Nilsson levels
or by assuming a weak octupole coupling, $\beta_3$ = 0.07~\cite{lea88}.
In summary, the low energy properties of the $^{223,225,227}$Ac isotopes have been
accounted for using the reflection asymmetric rotor-plus-quasiparticle
model~\cite{lea88} and, in this context, the $^{219,229}$Ac isotopes
have been described as shape transitional nuclei.

The study of transitional nuclei in the upper border of this octupole deformed region
is of relevance to understand the interplay of
octupole and quadrupole collectivities. In transitional nuclei the mechanism of grouping
rotational bands differs from that of parity-doublets in nuclei with stable octupole deformation.
Two intrinsic states (or rotational bands) with a given K and opposite parity have strong
octupole correlations, if the dominant one-quasiparticle component of the first partner
coupled to the octupole phonon of the core forms the largest collective component in
the structure of the second partner and vice versa. Experimentally one observes an enhancement
of the E1 strength for the parity pair partner bands.

With this aim the IS322 collaboration at ISOLDE-CERN has carried out a systematic investigation
of the heavy transitional Fr -- Th nuclei, for which scarce spectroscopic information was available.
These experiments have provided the first information on the absolute values of B(E1)
in this octupole transitional region. Relatively large B(E1) rates have been measured indicating
the presence of octupole correlations in $^{227}$Fr~\cite{kur97},
$^{229}$Ra~\cite{fra99} and $^{229}$Th~\cite{gul02,ruc06}.
Much weaker but still noticeable are the correlations
in the heavier $^{231}$Ra~\cite{fra01} and $^{231}$Th~\cite{aas99} isotopes.
We report here on the structure of their isobar, $^{231}$Ac.

Actinium-231 was first produced by bombarding $^{232}$Th with 14-MeV neutrons~\cite{cha73}.
Its half-life was then measured to be 7.5 $\pm $ 0.1 min and the strongest $\gamma$ rays
into $^{231}$Th were identified.
The proton pickup (t,$\alpha $) reaction was used to investigate the structure of
$^{233}$Pa and its isotone $^{231}$Ac~\cite{tho77}.
In the case of $^{231}$Ac the lack of information from beta decay caused that the band
assignments from the experimental cross sections were only
based on a comparison to those assigned to the isotone $^{233}$Pa.
A later study, and the only previous investigation where the $\beta^-$ decay of
\nuc{231}Ra has been reported~\cite{hil85}, assigned six gamma lines to $^{231}$Ac.
From their time behaviour a half-life of 103(3) s was obtained for \nuc{231}Ra,
but no attempt to build a level scheme was made.

In this paper the low-energy structure of $^{231}$Ac
fed in the $\beta^-$ decay of
$^{231}$Ra has been investigated by
$\gamma$-$\gamma$ and $\gamma$-e$^-$ spectroscopy.
Multipolarities of 28 transitions have been established
by measuring conversion electrons
with a \textsc{mini-orange} electron spectrometer.
The decay scheme of $^{231}$Ra~$\rightarrow$~$^{231}$Ac
has been constructed for the first time. The Advanced Time
Delayed $\beta\gamma\gamma(t)$ method~\cite{mac89} has been used to measure the
half-lives of several levels.
The deduced B(E1) rates are used to infer the possible presence of octupole correlations
in the $^{231}$Ac nucleus.

\section{Experimental Setup}

The study on  the \nuc{231}Ac structure is based on two sets of measurements
performed at the PSB-ISOLDE on-line mass separator \cite{kug92}.
In both experiments, a 1 GeV proton beam from the CERN PS-Booster
bombarded a UC$_2$-C target, producing via spallation reactions
the A=231 isobars.
In the first experiment the
excited structure of \nuc{231}Ac was populated directly from the
$\beta$ decay of \nuc{231}Ra with an estimated
production of 4$\times$10$^3$ atoms per $\mu$C. In the second experiment a Ta(Re) surface
ionizer was employed, favouring the ionization of Fr, and \nuc{231}Ac was populated through the
decay chain of \nuc{231}Fr $\rightarrow$ \nuc{231}Ra $\rightarrow$
\nuc{231}Ac \cite{fra01}, with an estimated yield of 7$\times$10$^4$ \nuc{231}Fr ions per $\mu$C.
The produced ions were accelerated to 60 kV and mass-separated before being collected onto a
magnetic tape transport system which connected two independent
but simultaneously operated measurement stations.

The radioactive beam was deposited in the centre of the
\emph{fast timing station} specially designed to measure the
half-life of excited states in the nano- and pico-second range by
the Advanced Time Delayed (ATD) $\beta\gamma\gamma$(t) method
\cite{mac89,mos89,mac91}.
The setup of this station consisted of a 3 mm thick NE111A plastic
scintillator positioned behind the tape at the collection point and
used as fast-timing $\beta$-detector. The thickness was chosen to
have an almost uniform response to the different $\beta$ energies.
A 2.5 cm thick BaF$_2$ scintillator was used as fast-timing gamma
detector. In addition, there were two HPGe detectors with relative efficiencies
of 70 \% and 40 \%, respectively, and energy ranges from 30 to 1500 keV.
They were used to provide a unique selection
of the $\gamma$ decay branch in the $\beta\gamma\gamma$(t) method and
to study $\gamma\gamma$ coincidences in order to determine the level scheme.
A detailed description of the setup is given elsewhere \cite{fra01}.
Up to seven parameters were recorded simultaneously for each event, namely the $\beta$ energy and
up to three $\gamma$ energies plus the time differences between the
$\beta$ signal and a given coincident $\gamma$ event.
The energy and efficiency calibrations of the gamma detectors were determined using a
\nuc{140}Ba source and an absolutely calibrated mixed source of
\nuc{57}Co, \nuc{88}Y, \nuc{109}Cd, \nuc{139}Ce, \nuc{113}Sn,
\nuc{137}Cs and \nuc{241}Am.

The A = 231 sample was transported one meter away by the tape to the \emph{conversion electron
station} specially designed for gamma identification and electron
conversion measurements. It consisted of an electron spectrometer,
a gamma telescope and a $\beta$ detector.

The electron spectrometer incorporated a \textsc{mini-orange} filter \cite{kli72,kli75}
and a 2 mm thick Si(Li) detector with an active area of 300 mm$^2$ and 1.9 keV energy
resolution at $\sim$100 keV. The detector was cooled to liquid nitrogen temperature by
means of a copper cold finger.
For the \textsc{mini-orange} filter a first set of permanent magnets was carefully
shaped according to the design of van Klinken and Wisshak. A relatively high transmission
of 1.8 \% of 4$\pi$ was achieved, but over a small energy range. Thus the set of magnets
finally chosen had a lower ($\simeq$0.5\%) 
but more uniform transmission over a broad range of energies
($\sim$25 $-$ 800 keV). This set consisted of three equally spaced rectangular magnetic plates
of 50 x 20 x 5 mm$^3$, with the largest dimension oriented radially. Measurements with different
\textsc{mini-orange} and detector positions showed that the transmission maximum was achieved
with de Si(Li) detector at 150 mm from the source and the \textsc{mini-orange} approximately midway
between the source and the detector. A central lead absorber was used to shield the detector from
direct X-rays and $\gamma$ rays.

The gamma telescope consisted of a 19.5 cm$^3$ planar HPGe followed by a 20 \% coaxial
HPGe, both in a single cryostat behind a 300 $\mu$m beryllium window.
The FWHM energy resolution varied from 0.4 keV at very low energies
up to 1.4 keV at E$_\gamma$=560 keV for the planar detector, and between 1.3
keV at E$_\gamma$=137 keV and 2.5 keV at E$_\gamma$=1836 keV for the coaxial
detector. Special care was given to detect low energy $\gamma$ rays with the
limit of sensitivity in the planar detector being 4 keV in single events and about 12 keV
in coincidence events. The Ge-escape peaks of the main L X-ray components of
Ra, Ac and Th from 4.9 to 6.7 keV are clearly identified. This blurs
unfortunately the possible contribution of any low energy gamma lines.
The L X-rays from \nuc{241}Am have allowed to extend the efficiency
calibration down to 13.4 keV. Moreover, $\beta$-gated $\gamma$ ray
singles spectra were also recorded with the coaxial Ge detector. For
beta-gating, a NE102 plastic scintillator was placed outside the
chamber in front of the gamma telescope. Simultaneously
to the gamma spectra, conversion electron singles spectra were
measured with the electron spectrometer. By dividing the acquisition
time into 8 consecutive subgroups, $\gamma$ multispectra from the planar Ge detector,
the $\beta$-gated coaxial Ge detector and the electron detectors were also
recorded allowing time evolution studies.
The e$^-$ and $\gamma$ signals were also recorded in
list-mode allowing for an identification of $\gamma$ lines in
coincidences down to 12 keV. The energy and efficiency calibrations for
these detectors were obtained using sources of \nuc{133}Ba and
\nuc{152}Eu, and the practically summing-free mixed source previously
described. Singles spectra from the planar detector as well as electron and $\beta$-gated
$\gamma$ singles spectra for the coaxial HPGe were collected in the
\emph{conversion electron station}.

Data were collected in three different modes of operation.
In the first two, measurements were conducted simultaneously
at the two stations, i.e. while a sample was measured in the \emph{conversion electron station}
a new one was collected at the \emph{fast timing} one. The time
cycle was chosen to enhance the nucleus activity of interest in the isobaric chain.
The time interval had to be slightly shorter than a multiple of the proton supercycle
from the CERN PS-Booster.
In the first experiment the cycles were chosen as 17 s and 190 s,
and in the second one as 26 s and 200 s, respectively, to enhance the
decay of \nuc{231}Fr (T$_{1/2}$ = 17.5(8) s) and \nuc{231}Ra (T$_{1/2}$ = 103(3) s).
In the third mode of operation the tape was stopped,
and the beam was continuously deposited at the collection point in order to
enhance the decay of \nuc{231}Ac to \nuc{231}Th (T$_{1/2}$ = 450 s).
The data acquisition was divided in 8 subgroups of equal time intervals
for gamma assignment and half-life determination.
In the first experiment about 7000 \nuc{231}Ra ions/s were transmitted to the
setup at the \emph{fast timing station}. The average count rate in the beta detector
was about 100 counts/s in the short cycle and 1350 counts/s in the long cycle.
Both cycles were used for the present analysis, with a total of 425 Mb of list mode data over
33 hours.
In the second experiment up to 110000 \nuc{231}Fr ions/s reached the \emph{fast timing station}.
The average count rate in the beta detector was $\sim$8000 counts/s in the short cycle.
In the long cycle the beam intensity was reduced to limit the count rate in the beta detector
to 10000 counts/s. Only data from the long cycle has been used for the present analysis,
with 5 hours of data taking.
The results on \nuc{231}Ra and \nuc{231}Th have been reported
in \cite{fra01} and \cite{aas99}, respectively.

\section{Experimental results}

In the present study, the results on the decay
of \nuc{231}Ra to \nuc{231}Ac are based on
the data collected with the long cycles. The intensity ratio of the
gamma lines between the short and long cycles has helped to
identify the low intensity gamma transitions. The most intense
$\gamma$ transitions have been assigned to the decay of \nuc{231}Ra by comparing
their temporal behaviour to that of the \nuc{231}Ac X-rays.
The fifteen most intense $\gamma$ transitions
have been used to deduce the half-life value.
The weighted mean value of the half-life is 104.1(8) s
in good agreement but more precise than the previous value~\cite{hil85}.
For more details of the analysis see~\cite{bou05}.

\subsection{$\gamma$ ray singles  and $\gamma $-$\gamma $ measurements}

\begin{figure}
  \begin{center}
\epsfig{file=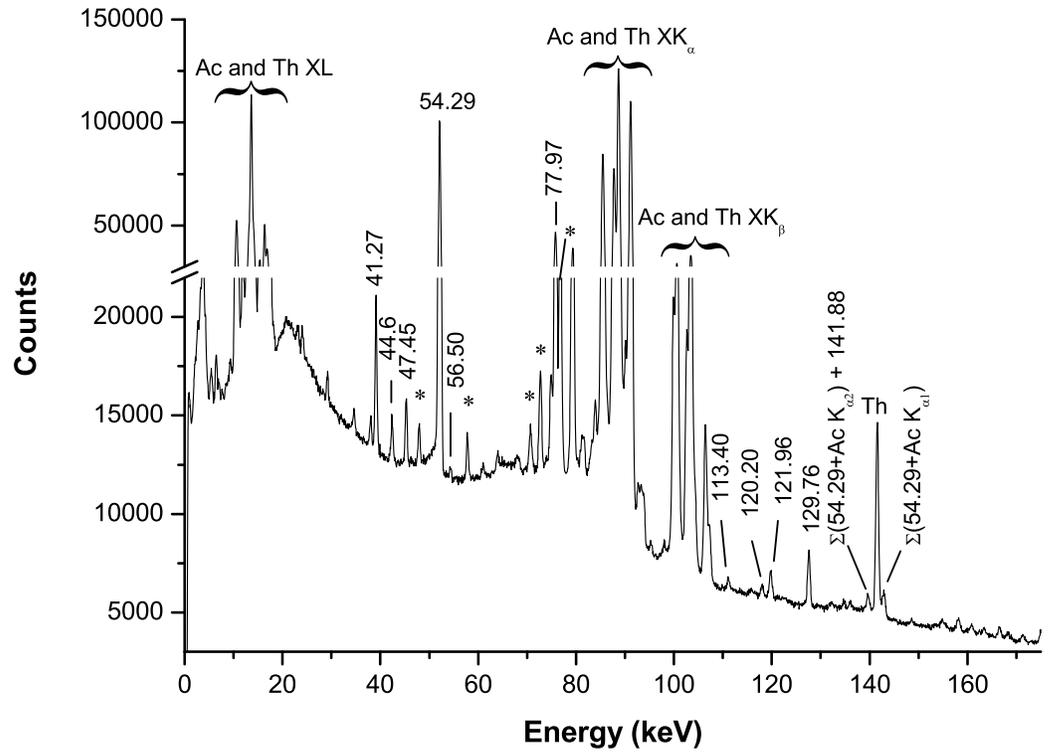, width=14cm}
  \end{center}
\begin{center}
\epsfig{file=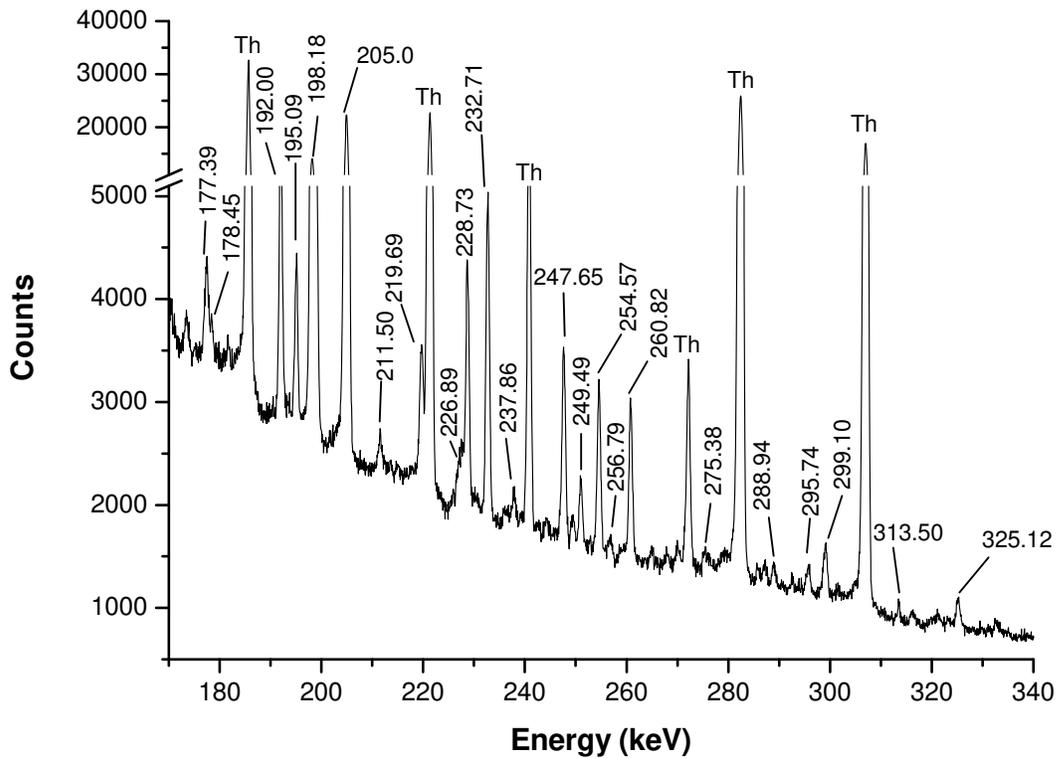, width=14cm} \caption{Low-energy part of the
gamma spectrum recorded in the planar Ge detector at the \emph{conversion electron station.}
The spectrum was taken during 4 hours and 52 minutes in the second experiment with a collection
and measuring time of 200 s.
The lines marked with an asterisk belong to the decay chain of the \nuc{212}Ra\nuc{19}F
contaminant.}
    \label{lowenergyplanar}
  \end{center}

\end{figure}

Fig.~\ref{lowenergyplanar} shows the low energy gamma spectrum recorded in the
planar detector at the \emph{conversion electron station}. Notice that
the sensitivity of the detector reached below 5 keV of energy. The aim was to identify the
expected low energy transitions as proposed to exist in
the 2p-2n neighbour $^{227}$Fr~\cite{kur97}.
The peaks observed between 4.8 and 6.4 keV have been identified as the K$_{\alpha}$ and
K$_{\beta}$ escape
 peaks in germanium of the most intense L X-ray components of Ac and Th.
These low energy peaks are about 10-12 \% of the L X-ray
assuming that the relative variation of efficiency
in this energy range is the same,
and that the efficiencies at these very low energies
followed the curves given in~\cite{han73}.
The lines marked with an asterisk belong to the decay chain
of the A=231 molecule $^{212}$RaF
observed mainly in the data of the second experiment.

The gamma transitions belonging to the $\beta$ decay of \nuc{231}Ac are
listed in  Table~\ref{gammastab}, the energies of the $\gamma$ transitions are the average of the
values obtained in the two experiments.
The uncertainties in the intensity values were calculated as a
weighted mean of the statistical errors and the error in the efficiency curve.
The latter one was estimated to be 10 \% for the planar detector
up to energies of 40 keV
and 6 \% for higher energies. The uncertainties of the efficiency curve of
the coaxial detector are estimated to be 5 \% up to 205 keV and 3 \% for higher energies.
Profiting from the good resolution of the planar
detector (0.4 keV at 6 keV and 0.5 keV at 54 keV)
some transitions could be only separated in this detector, in these cases the
intensity was taken from the planar detector only.
Summing of the K X-rays with the most intense $\gamma$ lines was observed
in the planar detector.
The summing contribution was estimated to be 1.2(2)~\% from the intensity of the peaks observed at
145.05(10) and 168.72(8) keV in comparison with the individual intensities of the K X-ray and the 54.29 and 77.97 keV $\gamma$ rays.
These low intensity peaks were not observed in the coaxial 20~\% HPGe detector
placed behind the planar detector.
Therefore we consider the relative intensity given in Table~\ref{gammastab}
for the 409.9 keV $\gamma$ transition as the one obtained
in the coaxial detector in order to avoid the possible contribution of summing of the 204.79(10)
and 205.00(10) keV $\gamma$ lines.
Most of the transitions have been observed in singles,
only some very low energy transitions have emerged in
e$^-$-$\gamma$ coincidences, see subsection 3.2. The relative
intensities of $\gamma$ transitions are listed in Table~\ref{gammastab}.
The 54.29 keV transition has been chosen
for the normalization, since the most intense one at 205 keV is a doublet
as clearly observed in the coincidence study.

Some $\gamma$ lines could only be observed or separated in coincidences and
therefore their relative intensity was obtained from coincidences
and marked with a footnote in Table~\ref{gammastab}.
The values obtained were compatible with the expected values in singles.
When the peaks are doublets with
$\gamma$ lines belonging to the decay of an isobar
or a contaminant, the extra contribution has been subtracted and
specifically indicated with a footnote.
In Table~\ref{gammastab}, the $\gamma$ transitions, their energies, intensities, the criteria used for the identification,
$\gamma$ transitions seen in coincidences and placement in the level scheme
are given.
A careful study of the shape of the 205 keV peak has been done
in the planar detector,
where the FWHM at this energy is of 0.7 keV. The least square fit to two peaks
indicates that their energies are at maximum 0.2 keV apart
with intensities of 40~\% and 60~\%
for peaks with centroids at 204.79 and 205.00 keV, respectively.
In the projection of the $\gamma$-$\gamma $ coincidence matrix a comparison of
intensity of the 205.00
and the 198.18 keV $\gamma$ lines both de-exciting the 266.76 keV state has given a
relative intensity for this component of 60\% of the total gamma intensity
in good agreement with the value deduced from singles.

\begin{figure}
  \begin{center}
\epsfig{file=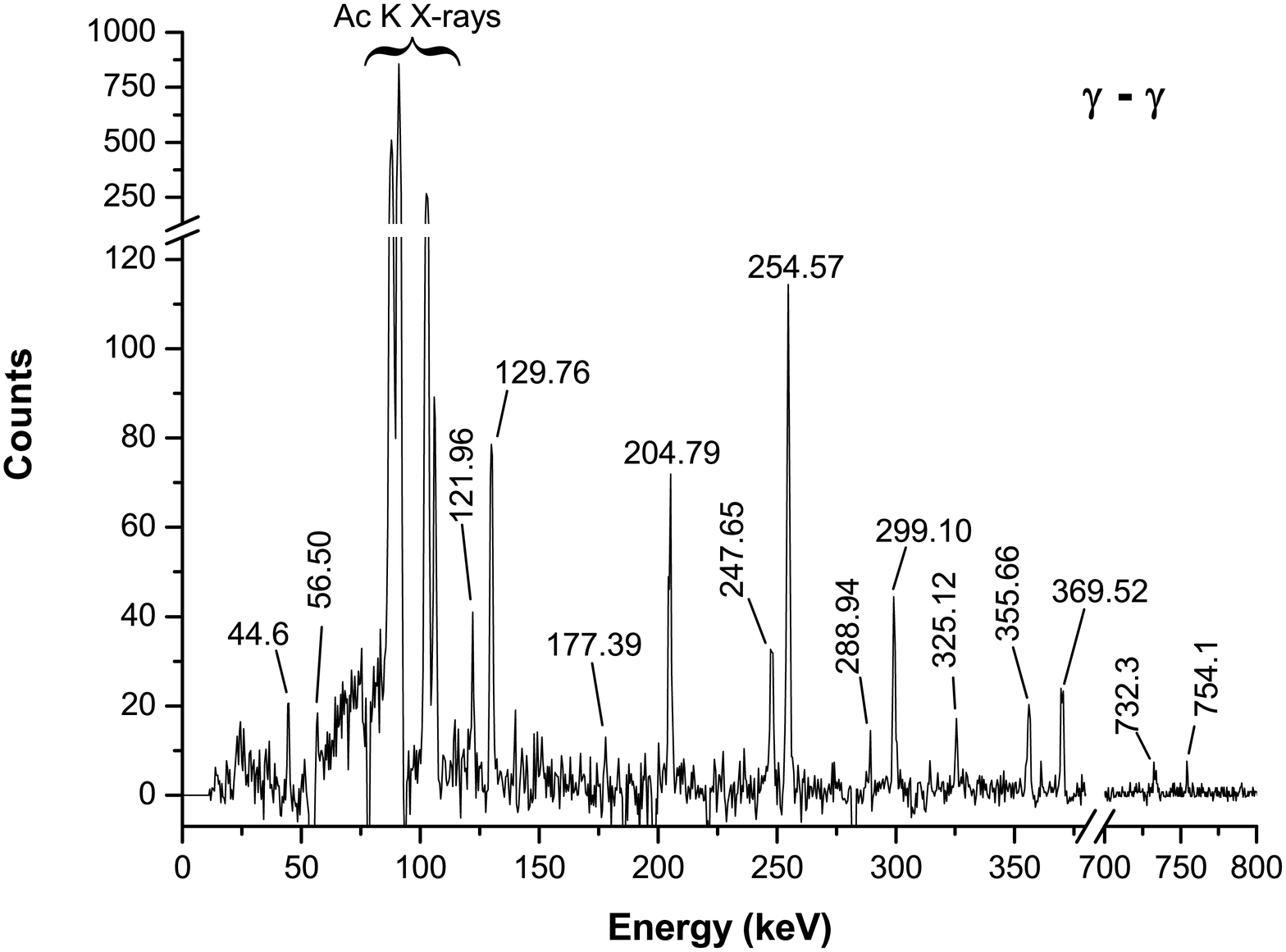, width=14cm}
\label{coine-54}
\epsfig{file=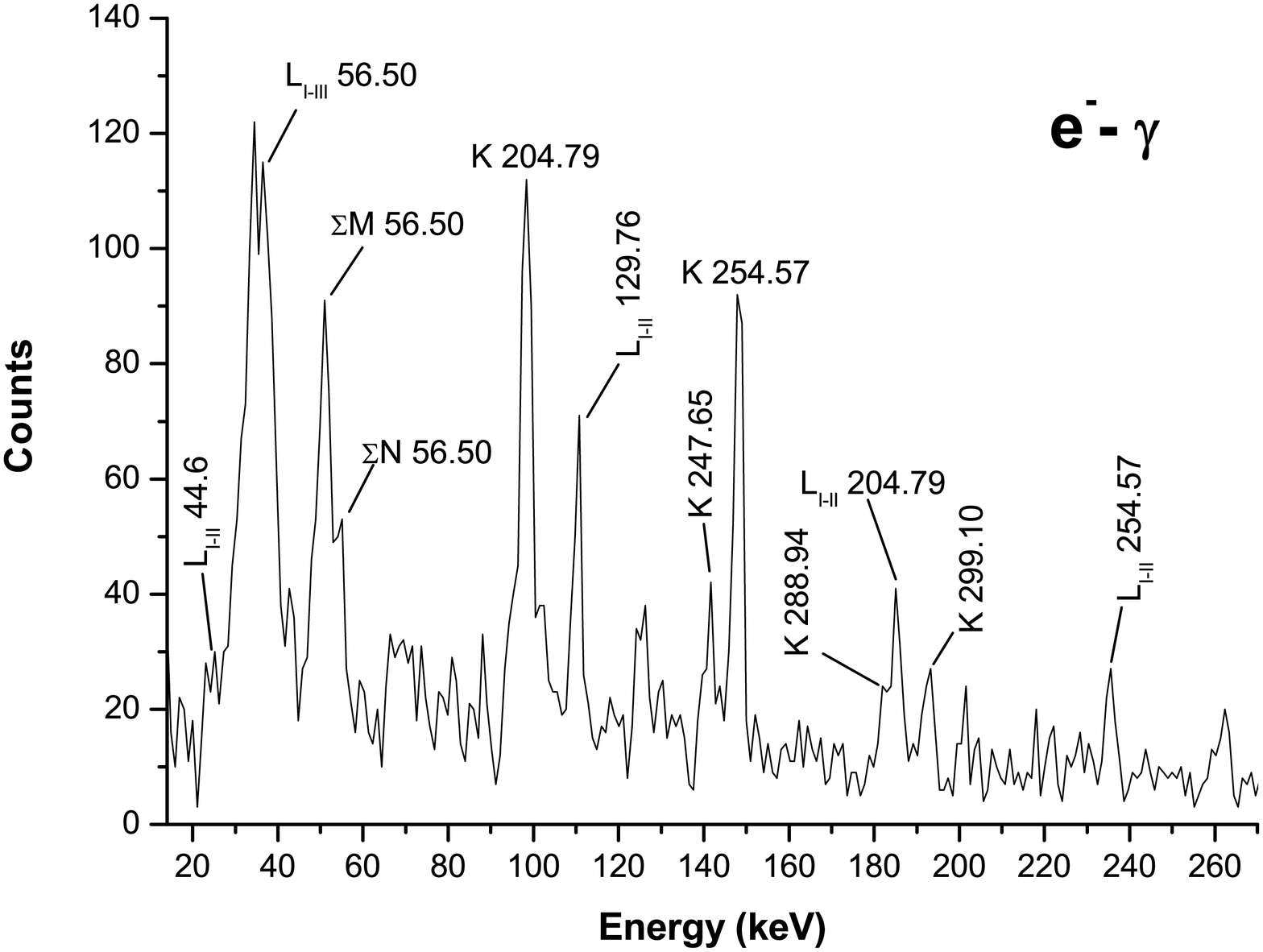, width=14cm}
\caption{Compton-subtracted projection spectra gated on the 54.29 keV $\gamma$ line.
In the upper part, the summed gamma coincidence spectrum gated on the HPGe
detectors from the \emph{fast timing station} is shown.
In the bottom part, the electron-$\gamma $ coincidence spectrum gated on the planar detector from the
\emph{conversion electron station} is displayed. The electron peaks are labelled with the energy values
of their corresponding $\gamma$ transitions.
The K, L$_{I,II}$, $\Sigma$M and $\Sigma $N electron binding
energies in Ac are 106.76, 19.57, 4.47 and 1.07 keV, respectively.
% Add binding energies
}
    \label{coin-54}
  \end{center}
\end{figure}

The $\beta$ gated $\gamma$-$\gamma$ coincidences were collected at the first
station and sorted off-line.
There are intense low energy transitions in $^{231}$Ac
coincident with many strong $\gamma$ lines, see, for instance,
the $\gamma$ spectrum coincident with the 54.29
keV transition displayed in the upper part of Fig.~\ref{coin-54}.
The spectrum shown is a sum of the
reciprocal projection placed in both HPGe
detectors of the first station.
As the threshold is different for these two detectors, the summed projection
is only reliable at energies higher than 30 keV.
For comparison the bottom part of Fig.~\ref{coin-54} shows the electron coincidence
spectrum gated on the 54.29 keV $\gamma$ line seen in the planar detector (for $\gamma$-e$^-$
coincidences see section 3.2).
Column 7 of Table~\ref{gammastab} lists the strong
$\gamma$-$\gamma$ coincidences observed in the present measurements; weaker
but still convincing coincidences are given within parenthesis.

The gate on the 198.18 keV $\gamma$ line shows in coincidence the Ac K X-rays and
the component of the 205-doublet at 204.79 keV. Therefore the $\alpha _{K}$ internal
conversion coefficient for this component is extracted from the ratio
of intensities of the K$_{\alpha}$ + K$_{\beta}$ and the 204.79 keV $\gamma$ line
corrected by the fluorescence yield $\omega_K$ for Ac equal to 0.969
(value taken from appendix F of~\cite{fir99}), following
the formula \(
\alpha _{K} = \frac{I_{K_{\alpha }} + I_{K_{\beta}}}{I_{\gamma }}\frac{1}{\omega_{K}} \label{eq:ICC}
\).
The resulting $\alpha _{K}$
value is 1.77(24) for the 204.79 keV $\gamma$ transition.

\subsection{Internal conversion coefficients and $\gamma$-e$^-$ coincidences}

\begin{figure}
  \begin{center}
\epsfig{file=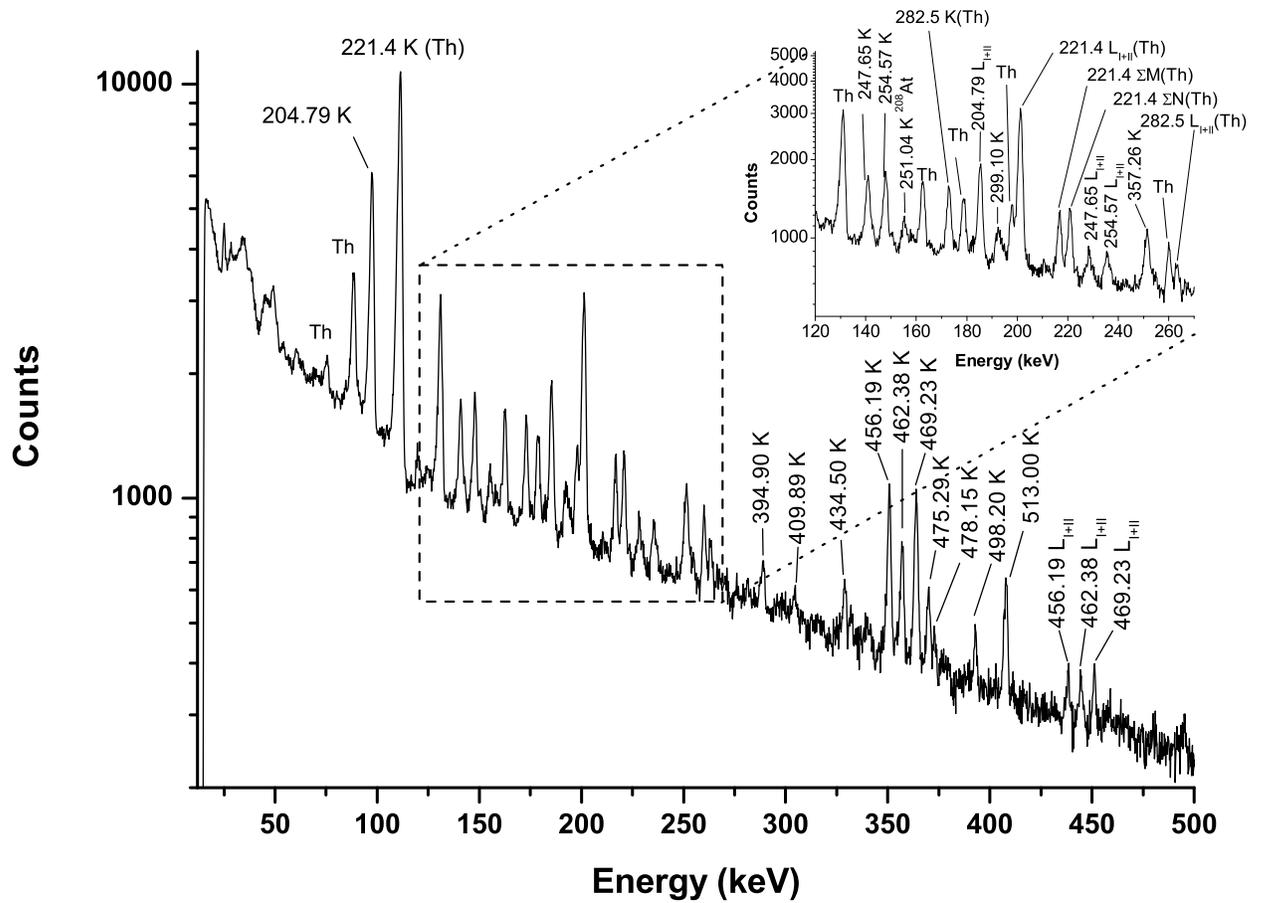, height=12cm}
\caption{Conversion electron spectrum taken with the long cycle of 200 s.
The electron peaks are labelled with the energy values of their corresponding $\gamma$ transitions.
The K, L$_{I,II}$, $\Sigma$M and $\Sigma $N electron binding
energies in Ac are 106.76, 19.57, 4.47 and 1.07 keV, respectively.
}
    \label{electrons}
  \end{center}
\end{figure}

The conversion coefficients were obtained from the simultaneous measurement
of gammas and conversion electrons at the second station.
The analysis was done using the data from the long cycle, where
the conversion electrons of the
66.2~keV transition in $^{231}$Ra that clearly dominate the electron
spectrum in the short cycle (see Fig. 4 of ref.~\cite{fra01}) are not visible.
The electron spectrum obtained
using the long cycle measurement time is shown in Fig.~\ref{electrons}.
The main electron lines correspond to the K components of the 221 keV M1
transition in $^{231}$Th and the 204.79 keV M1 transition in $^{231}$Ac.

The conversion coefficient values were obtained using the transmission function curve
of the \textsc{mini-orange} spectrometer, determined by internal calibration with
the conversion coefficients of the transitions in $^{231}$Th ~\cite{whi87}.
When an electron peak corresponds partially to conversion
electrons of $^{231}$Ac that overlap
with the ones of a line in \nuc{231}Th,
the latter contribution has been corrected for by
using the known conversion coefficients for transitions in \nuc{231}Th~\cite{whi87}.
The internal conversion
coefficients obtained in this work, the theoretical values from \cite{kib04}
and the multipolarities established for the transitions in \nuc{231}Ac are
presented in Table~\ref{convcoeftab}.
The total conversion coefficient, $\alpha _{T}$, needed to calculate
the intensity balance has been deduced from the experimental
conversion coefficient values in the following way.
If only one or two of the partial conversion coefficients are extracted from
the data and they are sufficient to determine the multipolarity (X$\lambda$) of the
transition, we deduce
$\alpha _{T}$(exp)= $\alpha _{i}$(exp)$\alpha^{th} _{T}(X\lambda)$/$\alpha^{th} _{i}(X\lambda )$
with i = K, L1+L2, $\Sigma $M or $\Sigma $N.
The total intensities obtained from the determined $\alpha _{T}$ values are given in the second column of Table~\ref{convcoeftab}.

The conversion coefficient for the doublet at 205 keV
has been obtained in the following way.
As previously explained we obtained an $\alpha _{K}$
value of 1.77(24) for the 204.79 keV transition from
$\gamma $-$\gamma $-coincidences corresponding to an M1 multipolarity.
This value  almost exhausts the full electron peak observed for the K component.
Knowing the M1 character of the 204.79 keV transition, we have used
the theoretical conversion coefficient to extract its contribution
to the electron spectrum and from that we estimate the contribution
of the 205.00 keV transition to the electron spectrum.
The extracted conversion coefficients given in
Table~\ref{convcoeftab} have large error bars, but clearly exclude the M1
multipolarity for the 205.00 keV transition.

The coincident events involving $\gamma$ rays in the planar Ge detector and
conversion electrons in the \textsc{mini-orange} spectrometer
have allowed us to identify
low energy transitions.
The e$^-$-$\gamma $ coincidences have also been used to identify to which nucleus
the converted gamma line belonged by studying
the coincident X-rays in the planar detector.

\subsection{Level lifetime measurements}

The Advanced Time Delayed (ATD) $\beta\gamma\gamma$(t)
technique detailed in \cite{mac89,mos89,mac91} has
been used to measure level lifetimes in \nuc{231}Ac.
This method involves triple coincidences of $\beta$-BaF$_2$-Ge events.
The timing information is provided by the coincidences between
the fast-timing $\beta$ and $\gamma$ (BaF$_2$) detectors,
while a coincidence with one of the two Ge detectors permits
to univocally select the desired decay branch. Two different
analysis methods are applicable, depending
on the lifetime range of the level of interest. Long lifetimes, observed as slopes
on the time-delayed tail of the time spectra, are measured by the
shape deconvolution method \cite{mac89}.
For lifetimes in the sub-nanosecond regime the centroid shift technique is used.
Both methods can be applicable in some cases.
Three calibration curves are essential for the centroid shift measurements.
The {\em prompt curve} describes the timing response of the BaF$_2$ detector as a function of the
energy of the full-energy peak from an incident $\gamma$ ray.
The {\em time response curve} provides the time response for the complete energy spectrum generated by a
monoenergetic transition; and, finally, the {\em Compton correction
curve} corresponds to the time shift between Compton events
of the same energy but originated by prompt primary $\gamma$ rays
of higher energies.
The first two curves were obtained off-line at ISOLDE using a
pre-calibrated source of \nuc{140}Ba $\rightarrow$ \nuc{140}La $\rightarrow$ \nuc{140}Ce
prepared at the OSIRIS fission product separator at Studsvik in Sweden \cite{mac95,fog97}.
The measurement of {\em the Compton correction curve} and a more exhaustive calibration of the
prompt curve were performed on-line at OSIRIS using a variety of beams while preserving,
as much as possible, the experimental geometry used at ISOLDE. Preliminary results of the lifetime values are given in~\cite{bou07}.

\begin{figure}
  \begin{center}
    \begin{tabular}{cc}
      \resizebox{60mm}{!}{\includegraphics{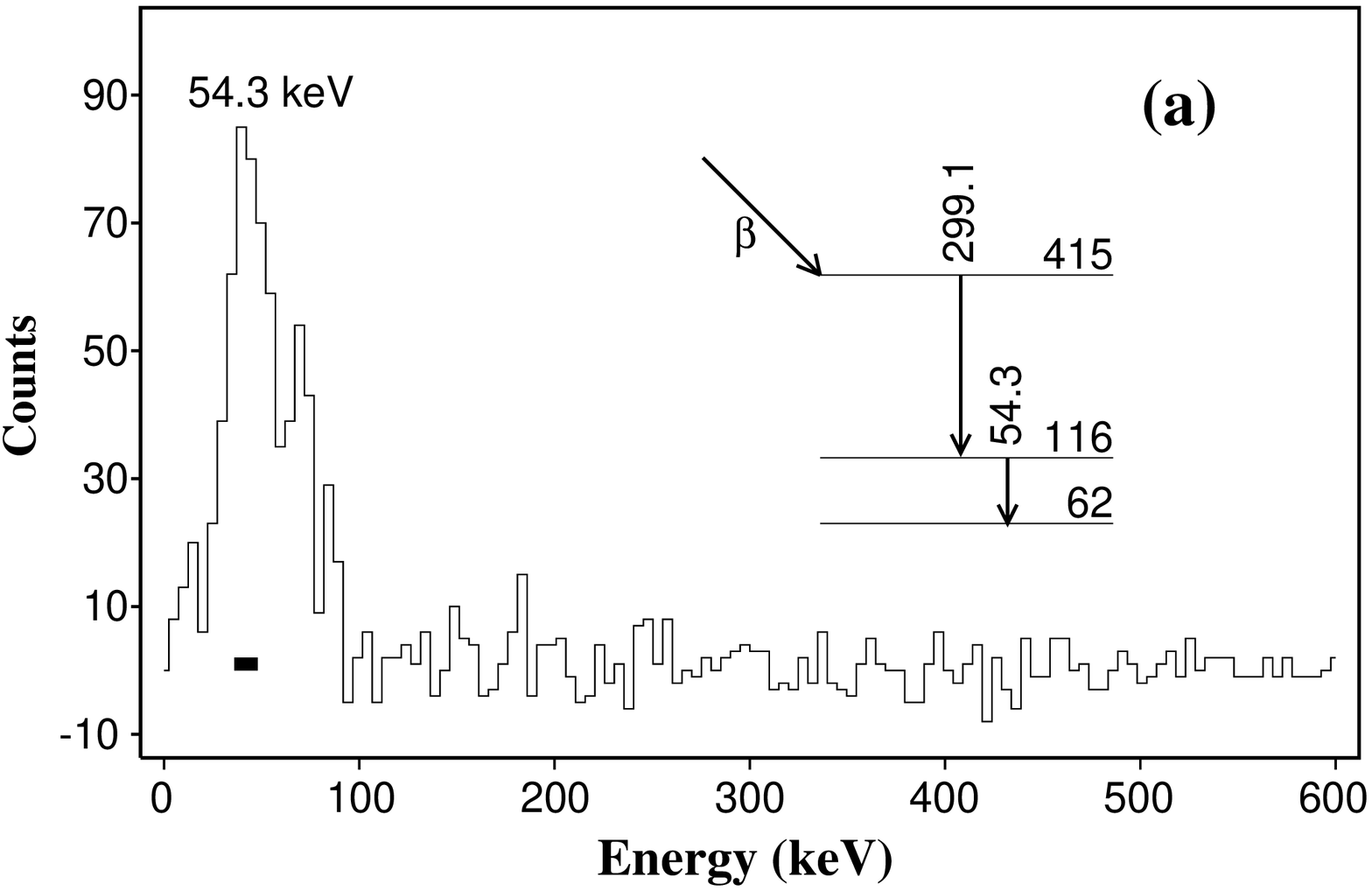}} &
      \resizebox{60mm}{!}{\includegraphics{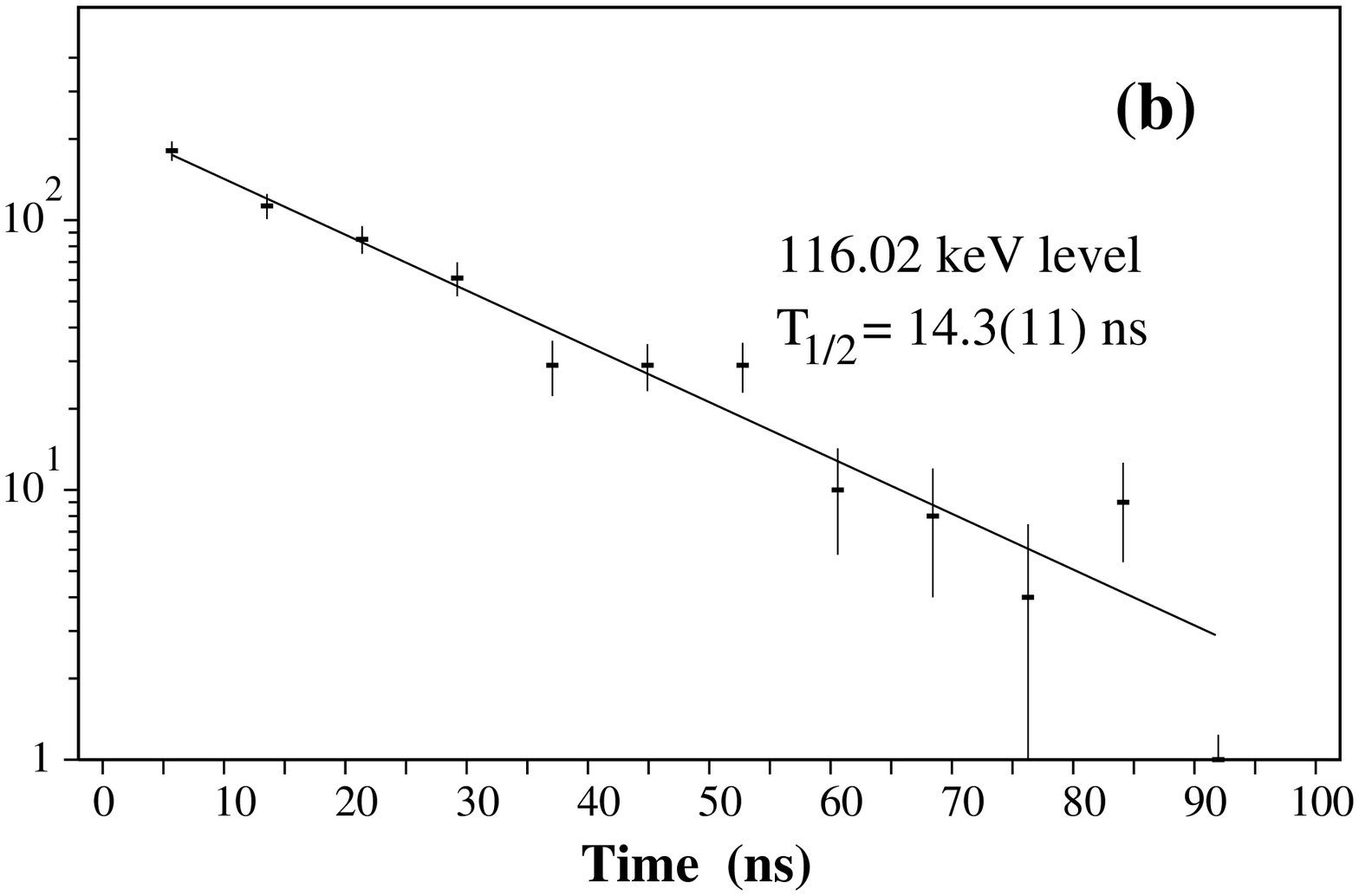}} \\
      \resizebox{60mm}{!}{\includegraphics{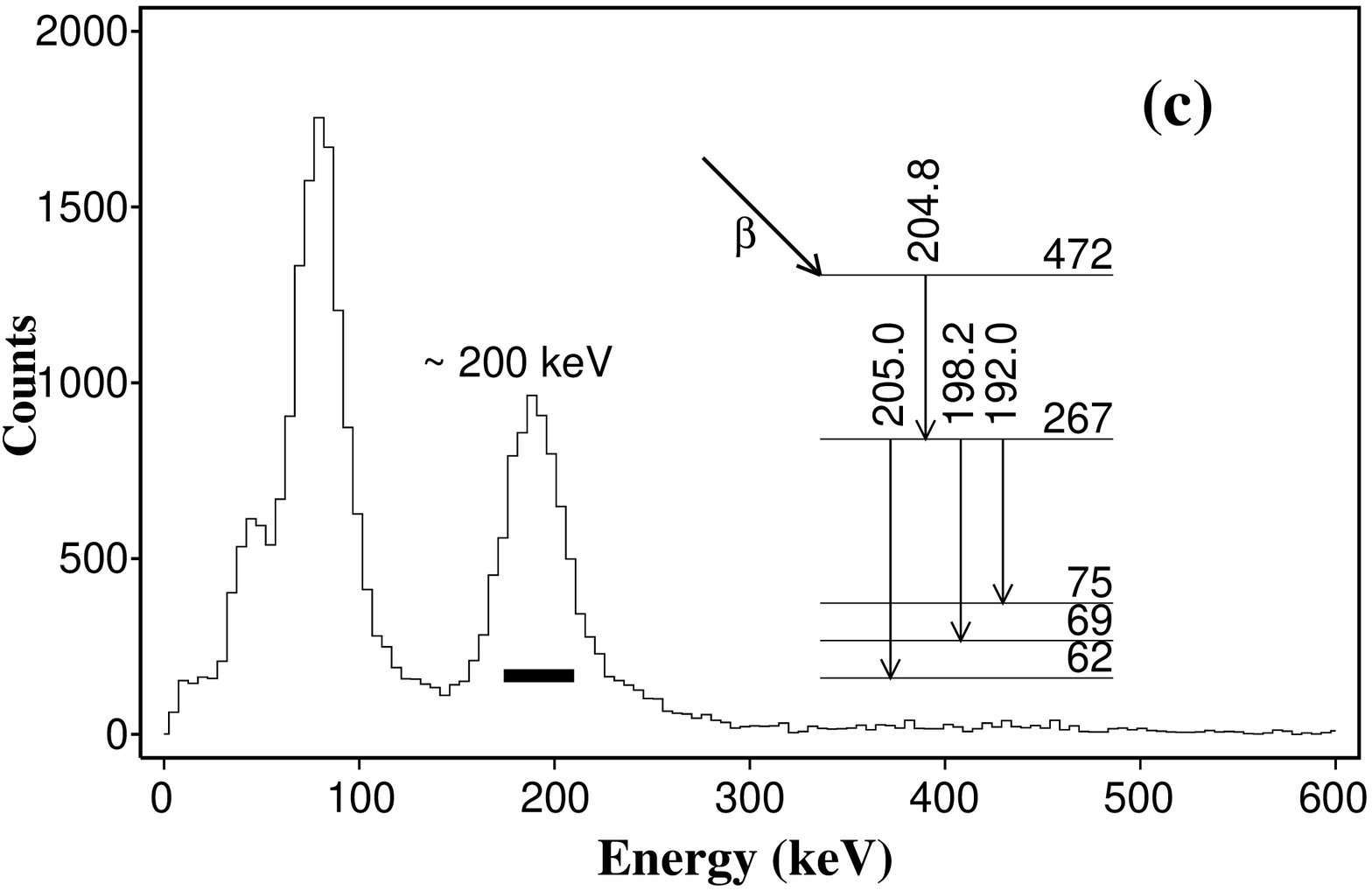}} &
      \resizebox{60mm}{!}{\includegraphics{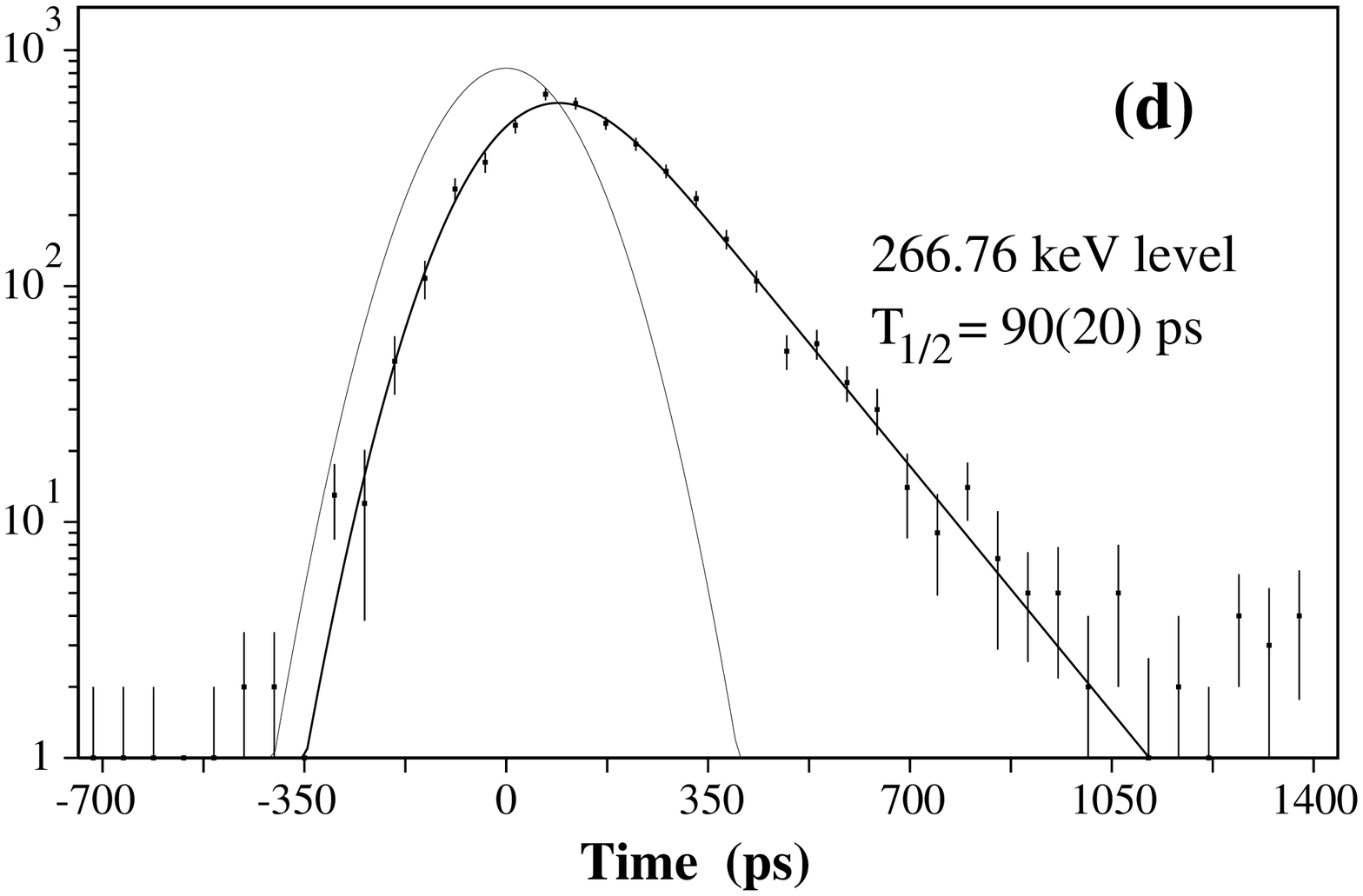}} \\
    \end{tabular}
    \caption{(a) and (c) BaF$_2$ energy
spectra sorted from the triple coincidence $\beta$-Ge-BaF$_2$ events.
The Ge-energy
gates were set on (a) the 299.10 keV and (c) the 205.0 keV transitions.
\newline
(b) and (d) Decay time curves
due to the lifetimes of the 116.02 and 266.76 keV levels, respectively.
These spectra
were obtained by selecting in the BaF$_2$ detector  the 54.29 keV $\gamma$ transition (marked by a black rectangle  in panel (a)) and the 200 keV region (marked in panel (c)).
 Panel (d) shows the experimental points, the prompt curve and the  slope curve fitted in the deconvolution method.}
    \label{timing}
  \end{center}
\end{figure}

\begin{itemize}
\item{Lifetime of the 116.02 keV level}\\
The slope-fitting method has been used to determine the lifetime of
the 116.02 keV level. A clear asymmetry is observed in the time
spectrum obtained as a projection onto the BaF$_2$-Ge fast time difference signal of the
54.3 keV transition selected in the BaF$_2$ detector and the 299.10 keV selected
in the Ge detectors. Due to the low statistics
of this time spectrum, it has been successively compressed in order to
increase the statistics and make use of a more convenient $\chi^2$ fit.
An average value of T$_{1/2}$ = 14.3(11) ns has been obtained.
The error bar takes into account the systematic error due to the use of a 
$\chi^2$ fit for a Poisson distribution 
and any possible deviation of the prompt time response from a Gaussian shape.
As an example we show in the panel (a) of Fig.~\ref{timing} the
BaF$_2$ energy spectrum projected from the $\beta$-Ge-BaF$_2$ data (the
gate in the Ge spectrum on the 299.10 keV line) and
in the panel (b) of
Fig.~\ref{timing} the slope fitting of the fast-timing
spectrum gated on the peak at 54.29 keV in the BaF$_2$ and compressed by a factor of 30.
\\
\item{Lifetime of the 160.73 keV level}\\
When gating on the 254.57 keV transition in the Ge detector spectrum and on the
54.29-77.97 keV group in the BaF$_2$ spectrum,
the centroid shift of the time spectrum is equal
to $\tau_0$ + $\tau_{415}$ + $\tau_{161}$ + $\tau_{116}$ where $\tau_0$ is the
prompt position, and $\tau_{415}$, $\tau_{161}$ and $\tau_{116}$ are the
mean lifetimes ($\tau$ = T$_{1/2}/\ln2$) of the 415.31, 160.73 and 116.02 keV levels, respectively. A reference
time spectrum obtained by gating on the 299.10 keV transition in the Ge
spectrum and on the 54.29-77.97 keV region in the BaF$_2$ spectrum has a
centroid shift  equal to $\tau_0$ + $\tau_{415}$ + $\tau_{116}$. The difference between
these two centroid shifts has been measured
to be $\tau_{161}<$ 1298 ps, which yields
T$_{1/2}$ $<$ 900 ps for the half-life of the 160.73 keV level.
\\
\item{Lifetime of the 238.01 keV level}\\
The double $\gamma$ ray cascade 247.65-232.71 keV was used for the determination
of the half-life of the intermediate level at 238.01 keV.
We observe a centroid shift between the projected $\beta$-BaF$_2$(t)
time spectrum obtained by selecting one transition in the BaF$_2$ detector and the other in the
Ge detectors and the one obtained by inverting the gates.
This centroid shift
directly provides the value of the lifetime of the 238.01 keV level. In this special case,
there is no need to know the prompt position as it cancels out when calculating the
difference between both centroid positions. The measurement gives a value of
T$_{1/2}$ = 57(11) ps for the 238.01 keV level.
\\
\item{Lifetime of the 266.76 keV level}\\
The two analysis methods, the deconvolution of the slope
and the centroid shift have been employed to measure
the half-life of the 266.76 keV level.\\
In the hereinafter discussion of the centroid shift analysis method, we adopt the following notation.
If we consider a cascade of two gamma lines $\gamma_1$ and $\gamma_2$, where the Ge
detectors are gated on the first transition $\gamma_1$, while a second gate on $\gamma_2$
is applied to the BaF$_2$ detector, the time spectrum obtained by the projection onto the fast TAC
will be labelled by S = [$\gamma_1$(Ge)-$\gamma_2$(BaF$_2$)].
For the application of the centroid shift method to the 266.76 keV level we consider on the one hand the
[198.18-204.8] cascade. The fitting of the centroid position in the time spectrum
yields $\tau_0$+$\tau_{472}$, where $\tau_0$ is the prompt response position and $\tau_{472}$ is
the mean lifetime of the 471.60 keV level, \( S_{1} = [198.18-204.8] = \tau _{0} + \tau_{472} \).
On the other hand, we take the [205.00-204.8] cascade. As we cannot resolve the two components
of the 205 keV doublet in the Ge detectors, the obtained time spectrum S$_2$ is actually the sum
of two theoretical time spectra S$_{21}$ = [205.00-204.8] and S$_{22}$ = [204.79-205.0] in the ideal
case where both components could be completely resolved in the detector. If $A$ and $B$ are the
fractions by which the S$_{21}$ and S$_{22}$ spectra contribute to the S$_2$ spectrum,
we will have:
\begin{eqnarray}
S_{2} = A \cdot S_{21} + B \cdot S_{22} \nonumber \\
S_{2} = A(\tau_0 + \tau_{472})+B(\tau_0 + \tau_{267} + \tau_{472}) \nonumber \\
S_{2} = \tau_0 +  \tau_{472} + B.\tau_{267} \nonumber
\end{eqnarray}
where $\tau_{267}$ is the mean lifetime of the 266.76 keV level. The difference of the two centroid shifts is then
\( S_{2} - S_{1} = B \cdot \tau_{267} \).
\\
The value of $B$ is estimated from the analysis of the sum of the 205.0 keV
gate projections onto the Ge detectors. $B$ is given by the ratio of the area
of the 205.00 keV singlet and the sum of the areas of the 192.00, 198.18 keV
peaks and the 205.0 keV doublet. This is so because the corresponding peak in
the BaF$_2$ detector includes all these contributions.
We deduce a value of 69(7) \% for B. This gives a value of T$_{1/2}$ = 89(11) ps for the half-life
of the 266.76 keV level.
\\
For the slope fitting analysis method, we had to generate a time-delayed
spectrum ``free" from the influence of the 471.60 keV level lifetime. In the same manner as for the
centroid shift, this has been achieved by subtracting the projected time spectra from the [205-205]
and [198-205] cascades, where the latter was corrected to account only for the contribution
of the 266.76 keV level lifetime. From the deconvolution and fitting of this time
spectrum, we obtain a value of T$_{1/2}$ = 85(29) ps as the average of the half-lives
from the last and more compressed spectra. This value is consistent with 89(11) ps
obtained from the centroid shift method. We adopt a value of T$_{1/2}$ = 90(20) ps for the half-life of the
266.76 keV level.
\\
\item{Lifetime of the 471.60 keV level}\\
The slope fitting of the time spectrum obtained when selecting the 198.18 keV line
in the Ge detector and the 204.79 keV transition in the BaF$_2$ de-exciting in cascade
the 471.60 keV level, allowed to obtain an upper limit for the
lifetime of this level. By averaging over various values obtained from the fitting of time spectra
with different compression factors, we obtain a value of 54 ps as a conservative upper limit for the
half-life of the 471.60 keV level.
\end{itemize}

\section{Decay scheme}
The level scheme of $^{231}$Ac adopted in this work
is shown in Figs. \ref{esq-1}, \ref{esq-2} and \ref{esq-3}.
The level scheme includes 38 excited states covering about half of
the Q$_{\beta}$ energy window, see Table~\ref{logfttab}.
A certain $\gamma$ transition is assigned to the decay if it fulfills
at least one of the following criteria: i) the time behaviour of
the intensity follows the
value of the $\beta$ decay half-life, ii) for the very low intensity
lines, the $\gamma$ line is assigned if the ratio of intensities
for the short and long cycle is the same as the one obtained for the
intense transitions already assigned, iii) it is seen in $\gamma$-$\gamma$ coincidences or iv)
it is seen in e$^-$-$\gamma$ coincidences with a transition already assigned.
In this way more than one hundred transitions have been placed in the level scheme, with
less than 2~\% of total intensity not being included in the level scheme.
For the majority of the transitions in the decay scheme the measured energy is within
0.35 keV of the energy differences of the initial and final state.
Total intensities were calculated using the determined
total conversion coefficient as explained in the subsection 3.2.
For the less intense transitions, whose conversion coefficients could not be
experimentally determined, the lowest conversion coefficient value compatible
with the parity of the two connecting states was adopted in most of the cases.
The levels assured by coincidences are shown in Figs. \ref{esq-1}, \ref{esq-2} and \ref{esq-3}
by continuous lines. The states determined by an energy fit are
considered if
at least two transitions connect the state to lower known states.
The levels determined by energetics are shown in Figs.~\ref{esq-1}, \ref{esq-2}, \ref{esq-3} with dashed lines.

The information available on the $^{231}$Ac levels prior to the present work
comes from the $^{232}$Th(t,$\alpha $)$^{231}$Ac reaction study~\cite{tho77}, where
17 levels were identified.
The ground state of $^{231}$Ac was assigned as the I = 1/2 member of the 1/2$^+$[400] band in
agreement with the predictions of the Nilsson model.
The level assignment was based
on the comparison
of the differential cross section estimates and experimental values.
The calculation of the differential cross section
used Nilsson wave functions with pairing and Coriolis effects
included. The deformation parameters $\varepsilon_2$ and $\varepsilon_4$ were treated as free parameters to reproduce the observed levels resulting in $\varepsilon_2$ = 0.24 and
$\varepsilon_4$ = -0.02.
The two main conclusions of the (t,$\alpha $) work~\cite{tho77}
relevant for our study are the assignments of the ground state as the 1/2$^+$[400] band head and
the 235(4) keV state as the 3/2$^+$[402] band head. Both states are observed in our work and
used to deduce the parity of the other
levels established in this $\beta$ decay work up to 600 keV in excitation energy.

Recently deep inelastic one-proton transfer studies on $^{137}$Cs produced by the $^{232}$Th+$^{136}$Xe reaction~\cite{bro99}
gave two extra bands tentatively assigned to $^{231}$Ac, since $^{231}$Ac is expected to be produced in this type of proton transfer reaction as the main reaction partner of $^{137}$Cs.
In this work Ac X-rays and several $\gamma$ lines, grouped in two bands with energies
between 100 and 500 keV, were observed. These bands were proposed to be the two-signature branches of a K=1/2
rotational band, as observed in $^{239}$Pu.
One of the bands was firmly proposed to be part of the ground state band of $^{231}$Ac with
the 163.3 keV transition assumed to connect to the ground state and as
such it has passed to the literature~\cite{bro99,sin05}. In our work we observe
a weak $\gamma$ line of 163.02 keV, but its relative intensity
with respect to the main $\gamma$ transitions in
$^{231}$Ac changes by a factor of two when the time of the cycle is changed. So it
was not assigned to the decay scheme. As none of the $\gamma$ lines of band A of
ref.~\cite{bro99} is observed in our work we conclude that they probably connect
high spin states not fed by $\beta$ decay.

We identify most of the low energy levels seen in the (t,$\alpha $) study although our
scheme is richer.
The starting point for parity assignment are the positive parity ground state
and the 3/2$^+$ state at 238.01 keV characterized in the reaction work~\cite{tho77}.
We assume that the $\beta$ feeding to the ground state is negligible,
since this is a $\Delta $I = 2 and $\Delta $K = 2
transition from the I$^{\pi }$ = 5/2$^+$[522] $^{231}$Ra ground state~\cite{fra01}.
For the log{\em ft}-value calculation~\cite{wil74} a Q$_{\beta }$ value of 2306(102) keV
is used together with the half-life value of 104.1(8) s obtained in this work.
The Q$_{\beta }$ value is derived from
the $^{231}$Ra mass measurement
($\Delta $M = 38226(19) keV)~\cite{her05} and the value given
in \cite{aud03} for the $^{231}$Ac mass ($\Delta $M = 35920(100) keV).
A summary of the beta feeding and $\log$ft values
for the different states is given in Table~\ref{logfttab}.
The $\beta$ feedings compatible with zero are not given.

The low energy part of the level scheme is based in the $\gamma $-$\gamma$ coincidence observed
 between the 37.8 keV and the 77.17 and 77.97 keV $\gamma$ doublet and the
54.29 keV and the 56.50 keV $\gamma$ transitions. The latter has also been observed in e$^-$-$\gamma$ coincidences.

The levels observed in this $\beta$ decay study are compared with the
quasiparticle states and energies obtained from a self-consistent
deformed Hartree-Fock calculation~\cite{vau73} with the Skyrme force
SLy4~\cite{cha98} and pairing correlations.
The BCS approximation was used with constant pairing gaps
for protons and neutrons of $\Delta_p$ = $\Delta_n$ = 0.8 MeV.
The self-consistent deformation in $^{231}$Ac is found at $\beta_2 = 0.22$
with $K^\pi=1/2^+$ as the ground state.
The deduced single particle energies place the band heads
$K^\pi=1/2^-$ at 192 keV($1/2^-$ [530]);
$K^\pi=3/2^+$ at 253 keV ($3/2^+$ [651]);
$K^\pi=3/2^-$ at 618 keV ($3/2^-$ [532]);
$K^\pi=5/2^-$ at 644 keV ($5/2^-$ [523]).
This calculation does not include any Coriolis mixing between
orbitals coming from the same shell, needed to explain
the low energy structure observed in $^{231}$Ac and other nuclei of the same region,
but it gives an idea of the ordering of the bands.
The excitation energy systematics of some of these bands in the neighbourhood
of $^{231}$Ac are shown in Fig.~\ref{bandas}. The band assignments for $^{231}$Ac
obtained in the reaction work~\cite{tho77}
are also displayed.
Although the knowledge on the excited structure of the isotones of Ac and Pa
in this region is reduced with little overlap, the comparison indicates that the systematics of excitation energy
for the $1/2^+$[400] band observed in \cite{bur03} can be extended
to the $3/2^+$[402] band. Furthermore, a parallel behaviour of the excitation
energy of the $5/2^+$[642] and $5/2^-$[523] bands is observed
for the isotopes of Ac and their isotones in Pa.

In the following the states established up to 550 keV excitation energy are discussed.
The more speculative level assignments are described at the end of the section.

{\em The ground state level}. The lowest level observed in the experiment
is assumed to be the ground state. It is fed by
low energy transitions such as the observed $\gamma$ lines of 18.44
and 37.8 keV and several transitions of
high energy and intensity that have not been observed
in coincidences, such as the 372.30, 498.20 and 513.00 keV
$\gamma$ lines. As discussed above, both theoretical
predictions and experimental findings
agree in the assignment of I$^{\pi}$ = 1/2$^+$ for the ground state.
In the (t,$\alpha $) reaction
the most populated level is the 1/2 member of the 1/2[400] band which
appears at zero energy~\cite{tho77}.
But as indicated by the authors, the high density of levels in
this mass region leads to a situation, where some peaks in the (t,$\alpha $) spectra
are unresolved multiplets, making the interpretation more difficult.
In fact the energy resolution in the (t,$\alpha$) work~\cite{tho77} was around 18 keV and
in our study several excited states have been identified within this energy interval.
Although the high density of
low-lying states in $^{231}$Ac does not allow to be conclusive about the assignment of the 1/2[400] band head as
 ground state, there is extra supporting evidence from the value of log$ft$ = 5.6, of the $\beta $ transition from
 the $^{231}$Ac ground state to the 555 keV level in $^{231}$Th. This
low log{\it ft} value can be explained, if
 we assume that the $^{231}$Ac ground state is the 1/2$^+$ 1/2[400] level decaying to the 1/2$^-$ 1/2[501] state as argued
in~\cite{egi81} for the case of $^{227}$Fr decay. The 1/2[400] Nilsson configuration originates from the s$_{1/2}$ proton orbit and
the 1/2[501] comes from the p$_{1/2}$ neutron orbit. This very favorable $\beta $
transition has been observed in several $\beta$ decays beyond $^{208}$Pb.
The excitation energy systematics of the 1/2[400] band is discussed in~\cite{bur03} for isotopes of Fr, Ac, Pa and Np.
These independent facts give us confidence in the
assignment of the 1/2$^+$[400] band head as the
ground state of $^{231}$Ac.

{\em The level at 5.6(4) keV} is established by coincidences. The parity is
 determined to be positive by the 232.71 keV E2(+M1) transition that de-excites the 3/2$^+$
238.01 keV state and corroborated by the rest of the transitions feeding the level.

{\em The level at 18.35(15) keV} is established by coincidences. The parity is assigned to be negative
due to the tentative E1 character of the 219.69 keV transition connecting the level to the
3/2$^+$ 238.01 keV state. We have identified the 18.44 keV
$\gamma$ transition
connecting this level to the ground state. Its multipolarity should be of E1 type, otherwise an unexpected extremely high $\beta$ feeding to this level needs to be considered. Thus this state is a
candidate for the band head of the 1/2[530] band. The 3/2$^-$
state of this band is the ground state of the isotone, $^{233}$Pa.

{\em The level at 37.95(15) keV} is fed by several $\gamma$ transitions and de-excited by the 19.64 and 37.8 keV
transitions. The latter has only been observed in
coincidence with the 77.17+77.97 keV  doublet, connected
through the 44.6 keV M1 transition.
In fact the intensity of the 37.8 keV transition given in Table 1
has been extracted from the comparison with the intensity of the 44.6 keV
transition observed simultaneously in the coincidence spectrum.
The 38 keV level was identified by Thompson et al.~\cite{tho77} as a doublet due to the
width of the observed peak (energy resolution $\simeq$ 18 keV) and the comparison of the experimental and the calculated
$^{232}$Th(t,$\alpha $)$^{231}$Ac differential cross section for the different excited states.
To explain the high value of the experimental differential cross section the contribution of the
I=3/2 member of the 1/2[530] band was assumed to be very close in energy.
In this work we have no indications of the double character
of the 38 keV state, as the 37.8 keV transition has
only been observed in coincidences with the 77.17+77.97 keV doublet.
Furthermore
we have not succeeded to place another level
nearby without great rearrangement of the level scheme.
The mixing of M1/E2 character of the 37.8 keV $\gamma$ line can balance
the in-out $\gamma $ intensity ($\alpha_T$ = 900(80)).

{\em The 61.73(20) keV level} is fed by very intense $\gamma$ transitions such as the 54.29,
205.00, 394.90, 409.89 and 469.23 keV transitions, and de-excited by the highly converted 56.50 keV
transition to the 5.6 keV state.
Notice that the 56.50 keV line has been observed
in coincidence with the 54.29 keV line both in $\gamma$-$\gamma$
and $\gamma $-e$^-$ coincidences, see Fig.~\ref{coin-54}, allowing to deduce the conversion coefficient
for the 56.50 keV transition given in Table~\ref{convcoeftab}.
The parity of this level is defined by the 54.29 E1 transition connecting it
to the negative parity state at 116.02 keV excitation energy.
As we will discuss later, we favour for this level a spin-parity
assignment of I$^{\pi }$ = 3/2$^+$.
One should notice that the band head of the 3/2[651] band
is assumed to be the ground state of the two-neutronless Ac isotope, $^{229}$Ac~\cite{tho77,bur03},
and is located at 86.48 keV in the isotone $^{233}$Pa (see Fig.~\ref{bandas}).

{\em The 68.57(20) keV level} is fed by more than ten transitions of E1 and M1 multipolarity.
The de-excitation occurs via the 63.23 keV $\gamma$ line
to the 5.6 keV state. A positive parity for this level is proposed based on the 198.18 keV E1
transition de-exciting the negative parity state at 266.76(10) keV.
The $5/2^+$ state of the $3/2^+$[651] is located 5 keV above
the band head in the $^{229}$Ac isotope and 8.2 keV in the
isotone $^{233}$Pa. This leads
to tentatively assign this level as the $5/2^+$ member of the $3/2^+$[651]
band in $^{231}$Ac.

{\em The 74.75 (20) keV level} is fed by several
transitions of 41.27, 192.00, (381.76), 456.19 and
1040.2 keV and de-excites
by the 36.74 keV transition to the  3/2$^+$ 37.95 keV state.
To fulfill the intensity balance a M1 multipolarity is needed for this transition
fixing a positive parity for the level.
Thompson and co-workers~\cite{tho77} identified a level at 76(5) keV
as the 9/2 member of the 3/2[651] band.
Due to the M1 transition to the 3/2$^+$ 37.95 keV state and the M1 character of the one coming from the
5/2$^+$ 531.00 keV state 3/2 or 5/2 spin values for this state are more plausible.

{\em The 116.02 (20) keV level}. This is a well established level populated by many transitions.
Its position is fixed by the coincidence between the 77.97 and 37.8 keV lines.
Taking into account the E1 character of the 77.97
keV transition feeding the 38 keV 3/2$^+$ state, the 116.02 keV state is given a negative parity.
There are two strong E1 transitions connecting this level to the 3/2$^+$ at 38 keV level and the tentatively assigned
3/2$^+$ state at 61.73 keV, so we assume
this level to be 1/2$^-$, 3/2$^-$ or 5/2$^-$.
The intensity of the eleven transitions feeding the state is balanced nicely by the four lines de-exciting it,
so no beta feeding is obtained, consistent with the spin-parity assignment of the level.
No equivalent level could be identified in the previous reaction work~\cite{tho77}.
A long half-life of 14.3(11) ns has been determined for this level
by the slope method, see section 3.3.
Similar half-life values for an excited state have been
measured in this region for the 3/2$^{+}$ 27.37 keV state (38.3(3)~ns)
in the isotope $^{227}$Ac \cite{bro01}, 5/2$^{+}$ 84.21 keV state (45.1(13)~ns)
in the isobar $^{231}$Pa \cite{bro01} and
36.5(4) ns~\cite{sin05} for the 5/2$^{+}$ 86.47 keV state in the isotone $^{233}$Pa.
The B(E1) value for the 27.37 keV transition in $^{227}$Ac connecting the
$3/2^+$3/2[651] $\rightarrow $ $3/2^-$3/2[532] states is similar~\cite{bro01} to the one obtained
for the 54.29 keV $\gamma$ transition in this study, supporting the assignment of this state as the
$3/2^-$3/2[532] band head.
This band was not observed in the work of Thompson et al.~\cite{tho77}.

{\em The 160.73 (15) keV level} is populated by the 77.17, (106.48), 254.57, 288.94, 295.74,
325.12 and 434.50 keV $\gamma$ transitions and de-excited by the 44.6 M1 transition
to the 116.02 keV state.
Due to the M1 character of the 44.6 keV transition we assign this level
as the 5/2$^-$ member of the 3/2[532] band based on the 116.02 keV state.
The half-life of the state, see section 3.3, has been determined to be less than 900 ps.

{\em The 238.01(15) keV level}. The position of the level is fixed by the coincidence between
the 121.96 keV and the 77.97 and 54.29 keV E1 transitions. The most intense transitions
de-exciting the state have been observed in coincidence
with the 177.39 and 247.65 keV $\gamma$ lines
placed above.
We assume that the level observed with high cross section in the (t,$\alpha $) reaction work~\cite{tho77}
at 235(4) keV corresponds to this level
observed in the $\beta$ decay of $^{231}$Ra.
The measured half-life of the 238.01 keV state is of 57(11) ps. No half-life of this order has been measured in other actinium isotopes. Equivalent values of half-lives have been measured in the two-proton
two-neutron neighbour $^{227}$Fr
for the 147 and 166 keV states identified as the 3/2$^+$ and 5/2$^+$
members of the 3/2[402] band ~\cite{kur97}. Therefore we identify the 238.01 keV level as
the 3/2$^+$ member of the 3/2[402] band (see Fig.~\ref{bandas}). This level has been observed
in $^{233}$Pa at 454.4 keV and at 336 keV in $^{229}$Ac. This is the lowest state for which an appreciable beta feeding is found from the in-out intensity balance.

{\em The 266.76(10) keV level}. The negative parity of this level is defined by the E1 character of the 205.00 and 198.18 keV
transitions to the 61.73(20) and 68.57(20) keV states of positive parity.
The decay pattern is very similar to the one of the 116.02 keV state.
The half-life of the level is 90(20) ps, see Fig.~\ref{timing} and section 3.3. Possible spin values for the level are 3/2$^-$ and 5/2$^-$. Due to the identification of a $\gamma$ transition to the ground state
a spin value of 3/2$^-$ is more favourable.
The beta feeding to this level is smaller than the error bar, so
no value is given in Table~\ref{logfttab}.

{\em The 415.31(15) keV level}. The position and parity of the level are assured by the coincidence between the 177.39 and the 232.71 keV transitions, the 254.57 M1 transition
with the 44.6 and 54.29 keV transitions, and the 299.10 M1 transition with the
77.97 and 54.29 keV E1 transitions.
The 116.02 and 160.73 keV states are of negative parity and are connected by M1
transitions of 299.10 and 254.57 keV to the 415.31 keV level. Therefore the latter
should also be of negative parity and spin value of either 3/2 or 5/2.
This level has a beta feeding of 5(2)~\%, giving a log{\em ft} of 6.7 (see Table~\ref{logfttab}).
There is a level identified in the reaction work~\cite{tho77} at 420(6) keV,
but no band assignment was given to this state.

{\em The 471.60(15) keV level}. This level is determined by the coincidence of the
204.79 keV M1 transition with the 205.00 keV $\gamma$ line and other transitions
de-exciting the negative parity 266.76 keV state. The negative parity assignment
is compatible with the non-observation of conversion electrons
of the relatively intense 403.03 keV transition that
connects this level to the 68.73 keV level of positive parity.
The very intense E1 $\gamma$ line of 409.89 keV has been assigned by an energy match
to connect this state with the 61.73 keV state.
The half-life of this level has been measured to be less than 54 ps.
A strong $\beta$ feeding of 27(10)~\% with log{\em ft} = 5.9 is obtained for this level.
This strong $\beta $ transition should
most probably connect the $^{231}$Ra ground state (5/2$^+$[622]) with a state of the same K.
Strong beta feeding up to 90~\% to levels of opposite parity but with the same K is rather common in this region,
see~\cite{fra01} for a systematics of $\beta $ transitions between the 1/2$^+$[400]$\longrightarrow $ 1/2$^-$[501] band heads. The compilation of log{\em ft} values~\cite{sin98} in
the actinide region (Z $>$ 82) lists two
first-forbidden transitions with $\Delta $I = 0 and $\Delta\pi $ = yes of the type 5/2$^+$ $\rightarrow $ 5/2$^-$ with
log{\em ft} value $\leq$ 6.
Thus we favour an assignment of 5/2$^-$ for this state.
This level could be the band head of the 5/2[523]
band observed in other isotopes of Ac, at 120.80 keV in $^{225}$Ac and at 273.14 keV in $^{227}$Ac, see Fig.~\ref{bandas}.
In our case the state is placed much higher in excitation energy, but
in our HF+Skyrme calculations the single particle state corresponding
to the 5/2$^-$[523] configuration moves from 219 keV for $\beta_2$ = 0.17 (with
this deformation the ground state is 1/2$^-$) to 597 keV for the case of
$\beta_2$ = 0.22 and 1/2$^+$[400] as ground state.
In the reaction work~\cite{tho77}
there is a level identified  at 469(10) keV with no tentative band assignment given.
Furthermore in our case we have identified three levels close to this excitation energy at 471.60, 473.40 and 478.30 keV.

{\em The 485.70(15) keV level} de-excites mainly by the following transitions: the 369.52 keV
transition in coincidence with the 54.29 and the 77.97 keV E1 transitions that
de-excite the negative parity 116.02(20) keV state,
the 247.65 keV M1 transition
in strong coincidence with the transitions de-exciting
the 238.01 keV 3/2$^+$ state and the 467.39 keV transition connecting the state to the 3/2$^+$ 37.95 keV state of positive parity.
Thus we assign the parity of the 485.70 keV state as positive.
Due to the M1 multipolarity of the transition to the 238.01 keV state
possible spin-parity for this level are: 1/2$^+$,
3/2$^+$ or 5/2$^+$. The log{\em ft} value of 6.7 excludes the possibility of a spin
value of 1/2$^+$.

{\em The 513.10(15) keV level} is determined by an energy match by the link of the
513.00 keV M1 transition to the ground state
and of the 494.57 keV transition to the 18.35 keV state.
A considerable $\beta$ feeding is found for this level, see Table~\ref{logfttab},
suggesting $\Delta $K = 0, $\pm$1.
The M1 transition to the ground state suggests I$^{\pi }$ = 1/2$^+$ or 3/2$^+$.
The relative strong beta feeding excludes the possibility of a spin value of 1/2$^+$
favouring 3/2$^+$ for this state.

{\em The 531.00(15) keV level} is determined by an energy match by the connection of
three of the most intense M1 transitions at 469.23, 462.38 and 456.19 keV to the
low energy levels at 61.73, 68.57 and 74.75 keV, respectively.
The parity of the level is defined positive and the most probable values for the spin are 3/2 or 5/2.
Considering the assignment of the parent $^{231}$Ra ground state as the 5/2$^+$[622]
and the strong beta feeding observed to this level (see Table~\ref{logfttab}) we assume the
531.00 keV level to be the 5/2 member of the 5/2[642] band.

{\em Other levels}\\
{\em A level
at 94(3) keV} was identified in the reaction work~\cite{tho77}. It was assumed to be the 7/2 member
of the 1/2[530] band. In our work we had identified a transition at 96.01 keV.
If this transition is the one de-exciting the proposed level, the spin of the state cannot be so high.
There is a weak $\gamma$ line of 141.88 keV that could feed this state from the 238.01 keV level and explain the coincidences between the 96.01 and 247.65 keV transitions. Energetically a transition from the 266.76 keV state
could be also identified. We tentatively include this state in the level scheme but not in the band assignment
shown in Fig~\ref{bandas-2}, as no M1 nor E2 transitions connecting the 96 keV level to the
18.35 keV state (E$_{\gamma }$ = 77.65 keV) have been identified.
One should note that there is an intense transition of 77.97 keV but of E1 character.\\
{\em The 245.78(20) keV level} is established by coincidences.
Two transitions in cascade of 21.0 and 129.76 keV connect the 116.02 and 266.76 keV levels of negative parity.
From the coincidence studies we could not decide the ordering, so the level could be placed
either at 137.00 or at 245.78 keV excitation energy.
The M1+E2 character of the 129.76 keV transition
gives negative parity to this new level.
The reaction studies~\cite{tho77} propose a level at 135(3) keV as the 13/2 member of the 3/2[651] band.
To avoid strong $\beta$ feeding to a level originally assigned as a high spin state, we
chose the placement of this state at 245.78 keV. This placement is more
consistent with the absence of $\gamma$ transitions to lower positive parity states.
It is difficult to understand how a 13/2 state could be populated from a 5/2$^+$ state in $\beta$ decay
as some $\beta$ feeding is assigned to this intermediate level.
\\
{\em The 372.30(10) keV level} has been tentatively placed in the level scheme connecting
the 372.27 keV E1 transition to the ground state and the 256.79 keV weak transition to the
116.02 state. Therefore this level would be of negative parity. No level at this energy has been
observed in the $^{232}$Th(t,$\alpha$) reaction work~\cite{tho77}.
Plausible spin assignments for this state are 1/2$^-$ or 3/2$^-$ and due to the
measured $\beta$ feeding we favour 3/2$^-$.\\
Other levels placed at 449.53, 478.30 and 498.20 keV excitation energies
have been established to account for strong $\gamma$ transitions
for which no coincidences have been found. In all cases several other weaker transitions
could be placed to strengthen the assignment.
For higher excited states the parity could not be determined.

A summary of the properties of the levels in $^{231}$Ac is given in Table~\ref{logfttab}. For each level
tentative spin and parity assignments as well as beta feeding, log{\em ft} values and half-lives are listed. Due to the density of low lying states and the absence of multipolarity determination for the low energy transitions, the beta feeding and  log{\em ft} values in Table~\ref{logfttab} should be taken with caution.

\section{Discussion}

M\"{o}ller et al. \cite{mol95} have tabulated the atomic mass excesses and
nuclear ground state deformation for most of the known nuclei.
The calculated Q$_{\beta}$ for $^{231}$Ra is 2.56 MeV in reasonable agreement
with the experimental value, 2.30(10) MeV.
They predicted for the
parent $^{231}$Ra and the daughter $^{231}$Ac
the same quadrupole deformation, $\beta _2$ = 0.207, and no octupole deformation for any of them.
Nevertheless moderately fast B(E1) rates connecting the lowest-lying
K$^{\pi }$ = 5/2$^{\pm }$ and K$^{\pi }$ = 1/2$^{\pm }$ bands were
found in $^{231}$Ra revealing the persistence of octupole collective effects~\cite{fra01}.
Therefore it is interesting to investigate
if some octupole collectivity is present in $^{231}$Ac.

\subsection{Rotational bands}

The spin of the parent, $^{231}$Ra, is assigned to be 5/2$^+$
in ref. \cite{fra01} and identified as the 5/2$^+$[622] Nilsson level, with quadrupole and
hexadecapole parameters $\beta_2$ = 0.17 and $\beta _4$ = 0.06.
The characterization of the $^{231}_{89}$Ac$_{142}$ ground state as the 1/2$^+$ member of the
1/2[400] band was done in
(t,$\alpha$) reaction studies~\cite{tho77}
and corroborated by the low log{\em ft} value of the $\beta $ transition to the 1/2[501] band head
in the daughter, $^{231}$Th~\cite{bro01}.
In the (t,$\alpha$) reaction work the structure of the heavy Pa and Ac isotopes were studied by the
U(t,$\alpha$)$^{233,235,237}$Pa and Th(t,$\alpha$)$^{229,231}$Ac reactions.
The comparison of Figs. 1 and 2 of their work indicates that the level structure and mainly the ground states
of the closest
isotone, $^{233}_{91}$Pa$_{142}$, and isotope $^{229}_{89}$Ac$_{140}$ are different from the one of $^{231}_{89}$Ac$_{142}$.
We could have
more similarities with the two-protonless isotone
$^{229}$Fr, but no information is available for this nucleus.
The
two-proton-two-neutronless neighbour, the
$^{227}_{87}$Fr$_{140}$ nucleus, has been studied by $\beta$ decay~\cite{kur97}
and the spins and parities
of the parent and the daughter are the same as in the case under study.

The band structure of $^{229}$Ac has been described fairly well both
considering quadrupole deformation plus Coriolis coupling as done in the reaction work~\cite{tho77}
or small octupole deformation~\cite{lea88}. The agreement was better when the octupole deformation term was included.
The lowest states were assumed to be part of the K = 3/2$^{\pm }$  and the K = 1/2$^{\pm }$ parity doublet bands.
A second K = 3/2$^{\pm }$ parity partners band is identified between the K= 3/2$^+$
at 336 keV and a second K= 3/2$^-$ which 3/2$^-$ band head was calculated to lie at 366 keV~\cite{jai90}. The band structure of the isotone, $^{233}$Pa, is explained considering
quadrupole and Coriolis coupling exclusively.

The ground state and the observed 38 keV state are assigned as the 1/2$^+$ and 3/2$^+$ members of the 1/2$^+$[400]
band based in the findings of the reaction work~\cite{tho77}.
Assuming that the moment of inertia
parameter $\hbar^2/2\Im$  has a value around 4-6 keV typical of this region, we expect the 5/2$^+$ member to be below the 3/2$^+$, between 15-34 keV excitation energy according to the relationship~\cite{boh95}
\begin{equation}
E_{I,rot} = \varepsilon _0 + \left(\frac{\hbar^2}{2\Im }\right)
\left[ I(I+1) + \delta_{K,\frac{1}{2}}(-1)^{I+\frac{1}{2}}(I+\frac{1}{2})a \right],
\label{eq:RotE}
\end{equation}

This is the case for the distribution of the excited states of the $1/2^+$[400]
band in the neighbours N= 140 $^{227}$Fr and $^{231}$Pa.
We have a level at 5.6 keV that could fulfill the requirements,
but a E2 transition connecting this state to the
ground state will exhaust the beta strength.
We are more tented to consider this level as belonging to the 1/2[660] band, due to
the strong Coriolis coupling of this band with the 3/2[651] band that could push the former down. In fact the intrinsic ground state of $^{231}$Ac is considered in~\cite{jai90} as a mixture of 1/2[400] and 1/2[660] bands.

The next candidates are the 74.75(68.57) keV states, where the zeroth-order position
of the band head ($\varepsilon _{0}$), the moment of inertia
parameter $\hbar^2/2\Im$ and the decoupling parameter $a$ are calculated to be
4.84(3.76) keV, 10.01(9.39) keV and 0.2667(0.3502) respectively.
The excitation energy for the $\frac{7}{2}^+$ state
is then calculated to be at 173.2(164.87) keV.
The deep inelastic 1p-transfer $^{232}$Th+$^{136}$Xe reaction work mentioned in the
 previous section~\cite{bro99} provides two extra bands tentatively assigned to $^{231}$Ac.
The first transition of 163.3 keV was assumed in the compilation~\cite{bro01} to connect a state at this energy
to the 1/2$^+$ ground state and therefore J$^{\pi }$=5/2$^+$ was inferred for this level.
In our work none of the proposed gamma rays have been observed,
but from this simple calculation it sounds plausible to place the observed
163 keV $\gamma$ transition on top of the
38 keV state, placing the $7/2^+$ member of the
1/2[400] band at 201 keV excitation energy.
This is compatible with the results in~\cite{bro99} as their setup was not
sensitive to low-energy gamma rays such as the 38 keV transition.

The 61.73 keV state is assigned as the band head of the 3/2[651] band
based on systematics. The 3/2$^+$ state of this band
is found at 40.09 keV in $^{225}$Ac, at 27.37 keV in $^{227}$Ac, is the ground state of $^{229}$Ac~\cite{tho77,bur03},
and is located at 94.66 keV in the isotone $^{233}$Pa.
The first excited member
of this band is located 5 keV above
the band head in $^{229}$Ac and 8 keV above in the isotone $^{233}$Pa. So we favour the assignment of
the 68.57 keV state as the first excited state of the band although the
74.75 keV state cannot be completely excluded.
The 9/2 and 13/2 members of this band were assigned in~\cite{tho77} as the states
at 76(5) and 135(3) keV excitation energy, respectively. As already discussed,
if the 76(5) keV state corresponds to the 74.75 keV state determined in this work,
it is very unlikely that the spin of the state is so high. Assuming that the
61.73 and the 68.57 keV states are the 3/2$^+$ and 5/2$^+$ members of
the band the 13/2 member is expected at 125 keV consistent with the findings
of the reaction work~\cite{tho77}.

We assume that the level placed at 238.01(15) keV is the 235(4) keV state
identified in the (t,$\alpha$) reaction work~\cite{tho77} and assigned
as the band head of the 3/2[402] band.
The excitation energy of this band has a behaviour with respect to
the neutron number similar to the one of the 1/2[400] band that becomes
the ground state in $^{231}$Ac
(see Fig.~\ref{bandas}).

The assignment of the 266.76 keV state is rather complicated. It is connected by very
intense E1 transitions to the 61.73 and 68.67 keV states assumed to be the band head
and the 5/2$^+$ members of the 3/2[651] band and with reasonable intensity to the
74.75 and the 37.95 keV states of the 1/2[400] band. The B(E1) values of the first two
transitions are a factor of 5 faster. So we favour the assignment of this state
as a member of a K= 3/2$^-$
band, see Fig.~\ref{bandas-2}. Similarly a K$^{\pi}$ = 3/2$^{-}$ band
was assigned in the two-protonless two neutronless $^{227}$Fr at 224.23 keV.

The 471.60 keV state is of negative parity due to the M1 transition to the 266.8 keV
 state.
 The strong $\beta$ feeding with log{\em ft} = 5.9 favours a
 first forbidden transition of the type 5/2$^+$ $\rightarrow $ 5/2$^-$. For more
 details see discussion in the previous section. Tentatively we assign this state as the
band head of the 5/2[523] band. This band constituted the ground state of $^{223}$Ac
 and it has been identified in $^{225}$Ac at 120.8 keV and at 273.14 keV in $^{227}$Ac.
The systematics shows a strong increase of the excitation
 energy with the neutron number in both Ac and Pa compatible with our placement at
 471.60 keV for $^{231}$Ac, see Fig.~\ref{bandas}.

Similarly strong $\beta$ feeding is found for the 531.00 keV state assigned as the
band head of the 5/2[642] band. This band is located
at 64 keV in $^{223}$Ac as the parity
doublet of the ground state band and it is also observed in $^{225}$Ac and $^{227}$Ac.
It is found in all Ac isotopes at 30-60 keV higher excitation energy than the 5/2[523] band,
in good agreement with the findings for $^{231}$Ac.

A summary of the tentative band assignments as discussed above
is given in Fig.~\ref{bandas-2}.

\subsection{Reduced transition probabilities}

The decay scheme shown in Figs. \ref{esq-1}, \ref{esq-2} and \ref{esq-3} has the
striking feature that several states at 116.02, 266.76 and 471.60 keV of
negative parity are connected through the most intense  E1 $\gamma$ lines to the
positive parity states at 61.73, 68.57 and 74.75 keV.
In this work we have directly measured the lifetimes of these levels and deduced the
B(E1) values for the de-exciting transitions.
The experimental B(E1) values together with their values in Weisskopf units are listed
in Table~\ref{redtransprob}. The B(E1) transition rates range
from 5$\times$10$^{-6}$ to 2$\times$10$^{-3}$ e$^2$fm$^2$.

The presence of octupole
phonon components enhances the E1 rates between members of parity partner bands.
Typical B(E1) values for enhanced E1 transition between K = 3/2$^{\pm }$ member
are, in this region, around 10$^{-4}$ e$^2$fm$^2$, whereas B(E1) values for nuclei
from a region without octupole collectivity are much
slower below 10$^{-5}$ e$^2$fm$^2$, see~\cite{ruc06,aas99}.
Strong mixing between bands is indicated by the fact that no
clear differences appear between B(E1) values for transitions
connecting the parity partner bands (intraband transitions) with
those connecting opposite parity bands with different K (interband transitions).
This effect has already been observed in $^{229}$Th~\cite{ruc06}.

In this work enhanced B(E1) rates in $^{231}$Ac are found for the
77.17 and the 121.96 keV transitions de-exciting the 238.01 keV state,
with B(E1) values of 4.4(12)$\times$10$^{-4}$
and 2.1(5)$\times$10$^{-4}$ e$^2$fm$^2$, respectively.
These transitions connect the 3/2$^+$[402] band with the band head and the 5/2$^+$
member of the 3/2[532] band. The ratio of B(E1) values, 2.1(8), is in good agreement with the theoretical intensity ratio given by Alaga's rule~\cite{alaga55} of 2.1(5),
giving support to  these assignments.
Similarly, the E1 transitions connecting
the 266.76(10) keV 3/2$^-$ state with the 61.73 keV 3/2$^+$ state and 68.57 keV
5/2$^+$ are moderately fast with B(E1)-values of 1.4(4)$\times$10$^{-4}$ and
1.4(3)$\times$10$^{-4}$ e$^2$fm$^2$, respectively, almost one order of magnitude faster than the
228.73 keV E1 transition to the 3/2$^+$ state of the 1/2[400] band, confirming the presence of octupole phonon components.

To ease the comparison of E1 strength over a wide range of nuclei, we use
the definition of the intrinsic dipole moment D$_0$, which removes the spin dependence of the B(E1) rates. Assuming a strong coupling limit and an
axial shape of the nucleus, D$_0$ is defined for K different from 1/2 via the rotational formula,
\begin{equation}
B(E1;I_i\longrightarrow I_f) = \frac{3}{4\pi}D^2_0<I_iK_i10\mid I_fK_f>^2
\label{eq:D-o}
\end{equation}

One should stress that the $|D_0|$ moment is used in this case as a convenient
parameter for inter-comparison, although this rotational formula
may not be strictly applicable for an octupole transitional nucleus.

The $|D_0|$ values have been calculated for the K$^\pi$=3/2$^\pm$ bands in \nuc{231}Ac,
for which experimental information exists. The values are compiled in Table~\ref{Do-moment}
and compared to those obtained for the nearby nuclei \nuc{227}Fr and \nuc{227}Ac. As stated before, the enhancement
of the B(E1) transition for K$^\pi$=3/2$^\pm$ partner bands is larger in the region than for other partner bands.
An average value of $|D_0|$= 0.044(4) e$\cdot$fm is obtained for the 3/2$^\pm$ with band heads at 116 and 238 keV, and
for the 3/2$^\pm$ bands at 266.8 and 61.7 keV we get $|D_0|$= 0.035(3) e$\cdot$fm.
These values are higher than those obtained for the lighter \nuc{227}Ac isotope and slightly lower than
the value for the 2.0 and 267.0 keV partner bands in \nuc{227}Fr.
Note that $|D_0|$ values of 0.077(3) e$\cdot$fm (on average) were obtained in \cite{ruc06}
 for the E1 transitions connecting the K$^\pi$=3/2$^\pm$ parity partner bands
in \nuc{229}Th with band heads at 0.0035 and 164.5 keV.

The effect of the enhancement of the E1 transition rates in \nuc{231}Ac seems to be still perceptible
and this may reveal the presence of weak octupole correlations in this transitional nucleus.

\section{Summary}
Our study has been focused on the properties of the low energy states in $^{231}$Ac
in a search for octupole correlations in this nucleus.
The level scheme of $^{231}$Ac has been determined from $\gamma $ and e$^-$ singles as well as $\gamma $-$\gamma $ and e$^-$-$\gamma $ coincidences.
More than 100 $\gamma$ lines have been assigned to the decay scheme.
The multipolarity of 28 transitions has been experimentally established. This has allowed to determine the parity of many levels. This study represents
the first decay scheme obtained for $^{231}$Ra and  enriches the known levels in $^{231}$Ac. We have identified up to a total of 30 new levels and assigned possible spin and parities for 21 of them.
The Advanced Time-Delayed $\beta \gamma \gamma(t)$ method was applied to measure lifetimes of several states in $^{231}$Ac.
Some particularly fast E1 transitions have been observed indicating that the octupole effects although weak still persist in this nucleus.

\section{Acknowledgements}
This work has been supported by the spanish CICYT, under projects FPA2002-04181-C04-02, FPA2005-02379 and FPA2007-62216 and the MEC Consolider project CSD2007-00042. We acknowledge the help and support of the ISOLDE Collaboration.

 %Tables

%1

\begin{table}[ht]
\tabcolsep2pt \caption{Gamma transitions belonging to \nuc{231}Ac. Tentative assignments are given in parenthesis. A tick in the S/L column indicates that the intensity ratio between the short cycle (17 or 26 s) and the corresponding long cycle (190 or 200 s) for a given transition is consistent with the $^{231}$Ac half-life.}
\label{gammastab}\vspace{0.5cm}
\newcommand{\minitab}[1][l]{\begin{tabular}{#1}\end{tabular}}
\begin{tabular}{ccccp{0.3cm}cccc}
\hline
\hline
E$_{\gamma}$(keV)&T$_{1/2}$&$\gamma$-$\gamma$&e$^-$-$\gamma$&S/L&I$_{\gamma}$&Coincidences&E$_i$(keV)&E$_f$(keV)\\
\hline\noalign{\smallskip}
18.44(10)&$\surd$&&$\surd$&$\surd$&~$\sim$ 90&&18.35&0.0\\
19.64(10)&&&&&$\sim$~ 60&&37.95&18.35\\
(21.0(4)&&&$\surd$&&$<7$&&266.76&245.78)\\
26.40(8)&&&$\surd$&$\surd$&$>$9(2)$^{a}$&&498.20&471.60\\
36.74(5)&$\surd$&&&&11(1)&&74.75&37.95\\
37.8(4)&&$\surd$&&&9(3)$^{b}$&&37.95&0.0\\
(40.30(5)&&&$\surd$&$\surd$&7(3)&&1155.30&1114.90)\\
41.27(5)&$\surd$&$\surd$&$\surd$&$\surd$&97(19)&129.8 205.0 247.7 254.6 &116.02&74.75\\
44.6(1)&$\surd$&$\surd$&$\surd$&$\surd$&30(4)&54.3 78.0 254.6&160.73&116.02\\
47.45(5)&$\surd$&$\surd$&$\surd$&$\surd$&38(5)&&116.02&68.57\\
54.29(5)&$\surd$&$\surd$&$\surd$&$\surd$&1000(112)$^{\dagger}$&44.6 56.5 77.2 122.0 129.8 (150.8)&116.02&61.73\\
&&&&&&177.4 204.8 247.7 254.6 288.9 299.1 &&\\
&&&&&&325.1 355.7 369.5 732.3 754.1 &&\\
&&&&&& 871.1&&\\
56.50(5)&&&&$\surd$&8(1)&&61.73&5.6\\
(63.23(5)&&&&&8(1)&&68.57&5.6)\\
70.44(5)&&&&$\surd$&8(1)&&485.70&415.31\\
77.17(7)&&$\surd$&&$\surd$&~~~~15(3)\multirow{2}{*}{\fontsize{30pt}{36pt}~~$\rbrace$}&37.8 (44.5) 77.2 122.0 129.8 205.0 &238.01&160.73\\
77.97(6)&$\surd$&$\surd$&$\surd$&$\surd$&378(44)&{247.7 254.6 288.9 299.1 355.7 357.3}&116.02&37.95\\
&&&&&& 369.5&&\\
81.48(9)&&$\surd$&$\surd$&&61(8)$^{c}$&&531.00&449.53\\
87.66(5)&$\surd$&&&&903(106)&&\multicolumn{2}{c}{Ac K$_{\alpha2}$ X-ray}\\
90.86(5)&$\surd$&&&&1411(165)&&\multicolumn{2}{c}{Ac K$_{\alpha1}$ X-ray}\\
96.01(6)&($\surd$)&$\surd$&&&30(7)&247.7 254.6&96.10&0.0\\
(106.48(9)&$\surd$&&&&50(8)$^{e}$&&266.76&160.73)\\
113.40(8)&$\surd$&($\surd$)&&$\surd$&11(1)&&485.70&372.30\\
\hline
\end{tabular}
\end{table}

\begin{table}[ht!]
Table 1, cont.\\
\tabcolsep2pt
\begin{tabular}{p{1.8cm}ccp{0.8cm}p{0.1cm}cccc}
\hline
\hline
E$_{\gamma}$(keV)&T$_{1/2}$&$\gamma$-$\gamma$&e$^-$-$\gamma$&S/L&I$_{\gamma}$&Coincidences&E$_i$(keV)&E$_f$(keV)\\
\hline\noalign{\smallskip}
120.20(7)&$\surd$&&&$\surd$&11(1)&(204.8)(e$\gamma$)&&\\
121.96(8)&$\surd$&$\surd$&$\surd$&$\surd$&28(3)&54.3 78.0 247.7&238.01&116.02\\
129.76(7)&$\surd$&$\surd$&$\surd$&$\surd$&59(7)&41.3 (47.5) 54.3 78.0 205.0&245.78&116.02\\
(134.38(10)&&&&$\surd$&9(2)&&372.30&238.01)\\
(141.88(10)$^i$&$\surd$&&$\surd$&$\surd$&$\leq$16(2)&&238.01&96.10)\\
150.75(10)&&($\surd$)&&$\surd$&8(1)&&266.76&116.02\\
(170.41(10)&&&&$\surd$&14(2)&&266.76&96.10)\\
177.39(8)&&$\surd$&$\surd$&$\surd$&~~~~19(3)\multirow{2}{*}{\fontsize{30pt}{36pt}$\rbrace$}&\multirow{2}{*}{54.3 78.0 232.7}&415.31&238.01\\
178.45(10)&$\surd$&&&$\surd$&15(2)&&&\\
192.00(8)&$\surd$&$\surd$&$\surd$&$\surd$&153(17)&205.0&266.76&74.75\\
195.09(10)&$\surd$&$\surd$&$\surd$&$\surd$&81(9)&(96.0) 247.7&680.80&485.70\\
198.18(8)&$\surd$&$\surd$&$\surd$&$\surd$&622(63)&205.0&266.76&68.57\\
204.79(10)&$\surd$&$\surd$&$\surd$&$\surd$&~~~483(164)\multirow{2}{*}{\fontsize{30pt}{36pt}$\rbrace$}&18.4(e$\gamma$) 21.0(e$\gamma$) 26.4(e$\gamma$) 37.8 &471.60&266.76\\
205.00(10)&$\surd$&$\surd$&$\surd$&$\surd$&728(159)&41.3(e$\gamma$) (44.5)(e$\gamma$) 54.3 78.0 81.5(e$\gamma$)&266.76&61.73\\
&&&&&&129.8 177.4(e$\gamma$) 192.0 198.2&&\\
&&&&&&205.0 (226.9)(e$\gamma$) 228.7 (237.9) 260.8&&\\
&&&&&&(871.1)&&\\
211.50(10)$^h$&&&&$\surd$&13(4)&&449.53&238.01\\
211.50(10)$^h$&&&&$\surd$&10(4)&&478.30&266.76\\
219.69(15)&$\surd$&$\surd$&$\surd$&$\surd$&91(12)&247.7&238.01&18.35\\
226.89(15)&$\surd$&&&$\surd$&25(9)&&&\\
228.73(10)&$\surd$&$\surd$&$\surd$&$\surd$&175(17)&205.0&266.76&37.95\\
232.71(9)&$\surd$&$\surd$&$\surd$&$\surd$&239(21)&177.4 247.7&238.01&5.6\\
237.86(15)&$\surd$&($\surd$)&&$\surd$&26(3)&&238.01&0.0\\
247.65(15)&$\surd$&$\surd$&$\surd$&&125(13)$^{d}$&18.4(e$\gamma$) (41.3)(e$\gamma$) (47.5)(e$\gamma$) 54.3 78.0 &485.70&238.01\\
&&&&&& (96.0) 122.0 195.1 219.7 232.7 &&\\
249.49(10)&$\surd$&&&$\surd$&27(4)&&&\\
\hline
\end{tabular}
\end{table}

\begin{table}[ht!]
Table 1, cont.\\
\tabcolsep2pt
\begin{tabular}{ccccp{0.3cm}cccc}
\hline
\hline
E$_{\gamma}$(keV)&T$_{1/2}$&$\gamma$-$\gamma$&e$^-$-$\gamma$&S/L&I$_{\gamma}$&Coincidences&E$_i$(keV)&E$_f$(keV)\\
\hline\noalign{\smallskip}
254.57(10)&$\surd$&$\surd$&$\surd$&$\surd$&165(16)&18.4(e$\gamma$) 41.3 44.5 47.5 54.3 &415.31&160.73\\
&&&&&& 78.0 96.0&&\\
256.79(15)&$\surd$&$\surd$&&$\surd$&22(3)&&&\\
260.82(10)&$\surd$&$\surd$&$\surd$&$\surd$&158(15)&204.8 &266.76&5.6\\275.38(10)&&&&$\surd$&16(6)&&1100.20&824.82\\
288.94(10)&$\surd$&($\surd$)&$\surd$&$\surd$&30(4)&&449.53&160.73\\
295.74(15)&$\surd$&&&$\surd$&19(5)&&456.56&160.73\\
299.10(15)&$\surd$&$\surd$&&$\surd$&66(7)&54.3 78.0&415.31&116.02\\
313.50(10)&&&&$\surd$&16(3)&&&\\
325.12(15)&$\surd$&$\surd$&&$\surd$&25(3)&&485.70&160.73\\
355.66(20)&$\surd$&$\surd$&&$\surd$&~~~~~~40(7)\multirow{2}{*}{\fontsize{30pt}{36pt}$\rbrace$}&&471.60&116.02\\
357.26(10)&$\surd$&$\surd$&&$\surd$&193(18)&\multirow{-2}{*}{54.3 78.0 }
&473.40&116.02\\
369.52(30)&&$\surd$&$\surd$&&148(26)$^{d}$&54.3 78.0&485.70&116.02\\
372.27(10)&$\surd$&$\surd$&&&455(43)$^{d}$& &372.30&0.0\\
(375.72(10)&&&&$\surd$&68(9)&&849.00&473.40)\\
(381.16(30)&&&&$\surd$&11(1)&&912.10&531.00)\\
(381.76(15)&&&&&30(5)&&456.56&74.75)\\
(387.99(15)&$\surd$&&&&45(12)$^{f}$&&456.56&68.57)\\
394.90(15)&$\surd$&&&$\surd$&127(13)&&456.56&61.73\\
396.92(15)&$\surd$&&&&57(18)$^{g}$&&415.31&18.35\\
403.03(15)&$\surd$&&&$\surd$&301(29)&&471.60&68.57\\
409.89(10)&$\surd$&$\surd$&$\surd$&$\surd$&1080(103)& &471.60&61.73\\
(417.55(10)&&&&$\surd$&14(2)&&485.70&68.57)\\
425.02(10)&$\surd$&&&$\surd$&23(5)&&670.80&245.78\\
429.62(15)&$\surd$&&&$\surd$&110(14)&&498.20&68.57\\
(432.00(30)&&&&$\surd$&36(5)&&847.40&415.31)\\
434.50(15)&$\surd$&&&$\surd$&143(13)&&595.15&160.73\\
\hline
\end{tabular}
\end{table}

\begin{table}[ht!]
Table 1, cont.\\
\tabcolsep2pt
\begin{tabular}{ccccp{0.3cm}cccc}
\hline
\hline
E$_{\gamma}$(keV)&T$_{1/2}$&$\gamma$-$\gamma$&e$^-$-$\gamma$&S/L&I$_{\gamma}$&Coincidences&E$_i$(keV)&E$_f$(keV)\\
\hline\noalign{\smallskip}
442.90(10)&$\surd$&&&$\surd$&43(6)&&680.80&238.01\\
(444.32(10)&($\surd$)&&&$\surd$&32(6)&&513.10&68.57)\\
445.74(10)&$\surd$&&&$\surd$&48(6)&&931.57&485.70\\
456.19(15)&$\surd$&$\surd$&&$\surd$&676(61)&&531.00&74.75\\
462.38(15)&$\surd$&$\surd$&&$\surd$&534(48)&&531.00&68.57\\
467.39(15)&$\surd$&$\surd$&&$\surd$&136(20)&&485.70&18.35\\
469.23(15)&$\surd$&$\surd$&&$\surd$&909(85)&&531.00&61.73\\
(473.40(30)&&&&$\surd$&28(6)&&473.40&0.0)\\
475.29(15)&$\surd$&$\surd$&&$\surd$&323(31)&&513.10&37.95\\
478.15(15)&$\surd$&&&&148(21)$^{f}$&&478.30&0.0\\
(481.74(30)&&&&$\surd$&25(5)&&931.57&449.53)\\
494.57(30)&($\surd$)&&&$\surd$&28(6)&&513.10&18.35\\
498.20(15)&$\surd$&&&$\surd$&227(21)&&498.20&0.0\\
513.00(15)&$\surd$&$\surd$&&$\surd$&659(86)$^{d}$&&513.10&0.0\\
569.4(5)&&&&$\surd$&51(7)&&1100.20&531.00\\
(577.7(3)&&&&$\surd$&19(6)&&1248.40&670.80)\\
586.8(6)&$\surd$&&&$\surd$&30(6)&&824.82&238.01\\
595.3(5)&&&&&51(7)&&595.15&0.0\\
607.6(5)&&&&$\surd$&10(1)&&&\\
(609.3(5)&&&&$\surd$&28(3)&&670.80&61.73)\\
(612.5(5)&&&&$\surd$&10(1)&&680.80&68.57)\\
(614.6(3)&&&&$\surd$&15(2)&&1100.20&485.70)\\
662.0(3)&&&&$\surd$&52(5)&&680.80&18.35\\
(666.3(4)&&&&$\surd$&17(3)&&1137.92&471.60)\\
\hline
\end{tabular}
\end{table}

\begin{table}[ht!]
Table 1, cont.\\
\tabcolsep2pt
\begin{tabular}{ccccp{0.3cm}cccc}
\hline
\hline
E$_{\gamma}$(keV)&T$_{1/2}$&$\gamma$-$\gamma$&e$^-$-$\gamma$&S/L&I$_{\gamma}$&Coincidences&E$_i$(keV)&E$_f$(keV)\\
\hline\noalign{\smallskip}
732.6(5)&$\surd$&$\surd$&&$\surd$&80(24)&&849.00&116.02\\
754.1(5)&&$\surd$&&$\surd$&31(5)&&870.16&116.02\\
763.1(3)&($\surd$)&&&$\surd$&45(12)&&824.82&61.73\\
842.0(5)&&&&$\surd$&18(6)&&847.40&5.6\\
844.2(5)&&&&$\surd$&18(3)&&&\\
849.1(5)&&&&$\surd$&13(5)&&849.00&0.0\\
857.8(6)&&&&$\surd$&35(6)&&&\\
868.4(6)&&&&$\surd$&69(11)&&1354.20&485.70\\
871.1(6)&&$\surd$&&$\surd$&60(6)&(205.0)&1137.92&266.76\\
912.1(6)&$\surd$&$\surd$&&$\surd$&150(26)&&912.10&0.0\\
937.7(5)&&&&$\surd$&32(9)&&&\\
986.9(4)&&&&$\surd$&31(9)&&&\\
1040.2(5)&&&&$\surd$&72(7)&&1114.90&74.75\\
(1046.2(5)&&&&$\surd$&27(3)&&1114.90&68.57)\\
(1086.3(6)&&&&$\surd$&24(3)&&1155.30&68.57)\\
(1150.1(4)&&&&$\surd$&31(4)&&1155.30&5.6)\\
1155.6(6)&&&&$\surd$&40(12)&&1155.30&0.0\\
(1248.3(5)&&&&$\surd$&7(2)&&1248.40&0.0)\\
(1354.4(9)&&&&$\surd$&10(2)&&1354.20&0.0)\\
\hline
\hline
\end{tabular}
\par
$^{\dagger}$ The absolute $\gamma $ intensity is obtained by multiplying by 7.58(19)$\times $10$^{-5}$
per decay (I$_{\gamma }$(54.29) = 0.0758(19) per decay). \\
$^{a}$ Intensity seen in singles 22(5), but there is a tabulated
26.0 keV Th line of relative intensity $<$0.006 of the 282 keV Th
$\gamma$ line~\cite{aas99}
which has been subtracted.\\
$^{b}$ Seen only in a coincidence spectrum (namely with 78.0 keV).\\
$^{c}$ Contribution from At X-ray subtracted.\\
$^{d}$ Contribution from Th subtracted.\\
$^{e}$ Contribution from Ac X-ray subtracted.\\
$^{f}$ Doublet. Contribution from Ra subtracted.\\
$^{g}$ Triplet with contribution of Ra (397.3 keV) and Th (396.9 keV).
The intensity corresponding to these components is subtracted.\\
$^h$ The intensity of this $\gamma$ line is shared between the two assignments.\\
$^i$ It could be due to summing of the 54.29 keV $\gamma$ ray and the 87.66 keV
Ac K$_{\alpha2}$ X-ray.\\
\end{table}
\clearpage

%2

\begin{table}[ht]
\caption{Experimental internal conversion coefficients for
transitions in \nuc{231}Ac. Theoretical values are taken
from \protect\cite{kib04}.}
\label{convcoeftab}\vspace{0.5cm} \tabcolsep5pt
\begin{tabular}{ccccccccp{1.9cm}}
\hline\hline
$E_{\gamma}$ & I$_{total}$& Initial & Conversion & \multicolumn{4}{c}{Internal
conversion coefficients} & Adopted  \\ \cline{5-8}
&& level & shell & ${\alpha}_{exp}$ & \multicolumn{3}{c}
{${\alpha}_{theo}$} &multipolarity  \\
\cline{5-8}
(keV)&&(keV)&&&E1&E2&M1&\\
\hline
44.6(1)&1100(300)&160.73&L1+L2&24(6)&0.467&192.2&27.30&M1\\
54.29(5)&1580(180)&116.02&L1+L2& $\leq $ 0.11&0.291&76.69&15.32&E1\\
56.50(5)&1200(200)&61.73&L1+L2&58(17)$^a$&0.264&62.72&13.63&E2(+M1)\\
	&&&$\Sigma$M&31(9)$^a$&0.096&31.38&3.292&\\
	&&&$\Sigma$N&12(4)$^a$&0.025&8.339&0.874&\\
77.17(7)&18(3)&238.01&L1+L2& $\leq $ 0.15&0.120&14.00&5.319&E1\\
77.97(6)&460(50)&116.02&L1+L2& $\leq $ 0.15&0.120&14.00&5.319&E1\\
129.76(7)&445(128)&245.78&L1+L2&1.31(40)$^a$&0.034&1.426&1.218&M1+E2\\
198.18(8)&670(70)&266.76&K&0.06(2)&0.075&0.163&1.964&E1\\
	&&&L1+L2&0.07(5)&0.012&0.242&0.366&\\
204.79(10)&1560(570)&471.60&K&1.77(24)$^{b}$&0.069&0.153&1.785&M1\\
205.00(10)&790(170) &266.76&K&$\leq $ 0.23$^{c}$&0.070&0.155&1.880& E1\\
	&&&L1+L2&$\leq $ 0.12$^{c}$&0.012&0.215&0.352&\\
232.71(9)&330(50)&238.01&K&0.12(6)&0.052&0.120&1.254&E2(+M1)\\
247.65(15)&290(40)&485.70&K&0.87(22)&0.045&0.106&1.054&M1(+E2)\\
&&&L1+L2&0.31(14)&0.007&0.101&0.196&\\
254.57(10)&365(60)&415.31&K&1.07(26)&0.042&0.100&0.977&M1\\
	&&&L1+L2&0.34(14)&0.007&0.091&0.181&\\
288.94(10)&60(13)&449.53&K&0.87(26)$^{a}$&0.032&0.078&0.688&M1\\
299.10(15)&130(30)&415.31&K&1.01(35)&0.029&0.072&0.626&M1\\
355.66(20)&64(14)&471.60&K&0.60(20)&0.020&0.051&0.385&M1\\
357.26(10)&310(50)&473.40&K&0.60(20)&0.020&0.051&0.385&M1\\
372.27(10)&465(45)&372.30&K&0.021(12)&0.018&0.047&0.344&E1\\
394.90(15)&170(30)&456.56&K&0.25(15)&0.016&0.041&0.293&M1\\
396.92(15)&77(25)&415.31&K&0.25(15)&0.016&0.041&0.293&M1\\

\hline
\end{tabular}
\end{table}

\begin{table}[ht]
Table 2, cont.\\
\vspace{0.5cm} \tabcolsep5pt
\begin{tabular}{ccccccccp{1.9cm}}
\hline\hline
$E_{\gamma}$ & I$_{total}$ & Initial & Conversion & \multicolumn{4}{c}{Internal
conversion coefficients} & Adopted \\ \cline{5-8}
&& level & shell & ${\alpha}_{exp}$ & \multicolumn{3}{c}
{${\alpha}_{theo}$} &multipolarity\\
\cline{5-8}
(keV)&&(keV)&&&E1&E2&M1&\\
\hline
409.89(10)&1100(100)&471.60&K&$<$0.035&0.015&0.038&0.265&E1\\
434.50(15)&180(25)&595.15&K&0.28(9)&0.013&0.034&0.227&M1\\
456.19(15)&840(90)&531.00&K&0.28(6)&0.012&0.031&0.199&M1\\
	&&&L1+L2&0.04(2)&0.002&0.012&0.036&\\
462.38(15)&660(70)&531.00&K&0.21(5)&0.012&0.030&0.192&M1\\
	&&&L1+L2&0.06(3)&0.002&0.012&0.035&\\
469.23(15)&1120(105)&531.00&K&0.21(5)&0.011&0.029&0.184&M1\\
	&&&L1+L2&0.04(2)&0.002&0.011&0.034&\\
475.29(15)&395(40)&513.10&K&0.21(6)&0.011&0.029&0.178&M1\\
478.15(15)&180(25)&478.30&K&0.15(6)&0.011&0.028&0.175&M1\\
498.20(15)&275(33)&498.20&K&0.17(7)&0.010&0.026&0.157&M1\\
513.00(15)&780(100)&513.10&K&0.14(4)&0.009&0.025&0.145&M1\\
\hline
\hline
\end{tabular}
\par
$^{a}$ Conversion coefficient deduced from $\gamma$-e$^-$ and $\gamma$-$\gamma$
coincidences with 54.29 keV $\gamma$ line.\\
$^{b}$ Conversion coefficient deduced from $\gamma$-$\gamma $ coincidences.\\
$^{c}$ Values extracted subtracting the contribution
of the calculated K and (L1+L2) electron conversion coefficients
of the 204.79 keV $\gamma$ line of
multipolarity M1.\\
\end{table}
\clearpage

%3

\begin{table}[ht]
\caption{Beta feeding (normalized to 100 decays of $^{231}$Ra)
and properties of the low spin levels in $^{231}$Ac}
\label{logfttab}\vspace{0.5cm} \tabcolsep5pt
\begin{tabular}{ccccc}
\hline
\hline
Energy (keV) &J$^{\pi}$&I$_{\beta}$(\%)$^a$ & log{\em ft} & T$_{1/2}$\\
\hline
       0.0 & $\frac{1}{2}^{+}$&           &                &             \\
    5.6(4) & $\frac{1}{2}^{+}$ ; $\frac{3}{2}^{+}$; $\frac{5}{2}^{+}$&           &                &             \\
 18.35(15) & $\frac{3}{2}^{-}$ &        &                &             \\
 37.95(15) & $\frac{3}{2}^{+}$&           &                &    \\
 61.73(20) & $\frac{3}{2}^{+}$&           &                &             \\
 68.57(20) & $\frac{5}{2}^{+}$ &           &                &             \\
 74.75(20) & $\frac{3}{2}^{+}$; $\frac{5}{2}^{+}$&           &    &             \\
96.10(20) & ($\frac{1}{2}^{-}$; $\frac{3}{2}^{-}$)&           &    &             \\
116.02(20) & $\frac{3}{2}^{-}$&   &              & 14.3(11) ns \\
160.73(15) & $\frac{5}{2}^{-}$&    &      &  $<$ 900 ps \\
238.01(15) & $\frac{3}{2}^{+}$& 1.2(7)  & 7.5       &   57(11) ps \\
245.78(20) & ($\frac{1}{2}$; $\frac{3}{2}$; $\frac{3}{2}$)$^{-}$ &   & &     \\
266.76(10) & $\frac{3}{2}^{-}$ &   &        &  90(20) ps \\
372.30(10) & ($\frac{1}{2}^{-}$); $\frac{3}{2}^{-}$&4.2(14)  &  &     \\
415.31(15) & $\frac{3}{2}^{-}$; $\frac{5}{2}^{-}$&5(2)   & 6.7       &       \\
449.53(30) & $\frac{3}{2}^{-}$; ($\frac{5}{2}^{-}$)&    &   &          \\
456.56(15) & $\frac{3}{2}^{+}$; $\frac{5}{2}^{+}$ & 2.5(8)   & 6.9   &       \\
471.60(15) & $\frac{5}{2}^{-}$ & 27(10)   & 5.9      &  $<$ 54 ps  \\
473.40(15) & $\frac{1}{2}^{-}$; $\frac{3}{2}^{-}$& 2.2(8)      &   7.0      &             \\
478.30(10) & $\frac{1}{2}^{+}$; $\frac{3}{2}^{+}$ &1.7(6)   & 7.1       &             \\
485.70(15) & $\frac{3}{2}^{+}$; $\frac{5}{2}^{+}$&4.5(15)& 6.7 & \\
498.20(20) & $\frac{3}{2}^{+}$& 3.9(13)   & 6.7      &   \\
513.10(15) & $\frac{3}{2}^{+}$&11(4)    &       &     \\
531.00(15) &  $\frac{5}{2}^{+}$	& 24(8)	& 5.9  &	\\
595.15(20) & $\frac{3}{2}^{-}$&2.1(7)   & 6.9       &             \\
670.80(20) &	&0.30(12)&   7.7  &	\\
680.80(30) & &1.8(6)   & 6.9      &             \\
\hline
\hline
\end{tabular}
\end{table}

\begin{table}[ht]
Table 3, cont.\\
\vspace{0.5cm} \tabcolsep5pt
\begin{tabular}{cccccc}
\hline
\hline
Energy (keV) &J$^{\pi}$&I$_{\beta}$(\%)$^a$ & log{\em ft} &T$_{1/2}$\\
\hline
824.82(30) & &0.5(2)   & 7.3       &             \\
847.40(30) & &0.5(2)   &  7.3    &   \\
849.00(20) & &1.5(5)   & 6.8      &             \\
870.16(40) & &0.3(1)    & 7.5     &             \\
912.10(20) & 	&1.5(5) 	& 6.7&	\\
931.57(20) & &0.7(2)   & 7.1       &             \\
1100.20(30) &	&0.8(3)	& 6.8 &	\\
1114.90(40)& &0.8(3)   & 6.8       &             \\
1137.92(40)& &0.7(2)    & 6.8      &             \\
1155.30(30)& &1.0(3)   & 6.6      &             \\
1248.40(50)& &0.24(9)   & 7.1      &             \\
1354.20(50)& &0.7(2)   & 6.5      &             \\
\hline
\hline
\end{tabular}
\par
$^{a}$ The $\beta$ feeding  is considered only when it is larger than the error bar.\\
\end{table}
\clearpage

\begin{table}[ht]
\caption{Experimental reduced transition probabilities in \nuc{231}Ac. }
\label{redtransprob}\vspace{0.5cm} \tabcolsep5pt
\begin{tabular}{cccccc}
\hline
\hline
Level(keV)&T$_{1/2}$&E$_{\gamma}$ (keV)&X$\lambda$&B(X$\lambda$)&B(X$\lambda$) (W.u.) \\
\hline
116.02(20)&14.3(11) ns&41.27&E1$^a$&1.8(4)$\times$10$^{-5}$ e$^2$fm$^2$&7.4(16)$\times$10$^{-6}$\\
	&&47.45&E1$^a$&4.6(8)$\times$10$^{-6}$ e$^2$fm$^2$&1.9(3)$\times$10$^{-6}$\\
	&&54.29&E1&8.2(7)$\times$10$^{-5}$ e$^2$fm$^2$&3.4(3)$\times$10$^{-5}$\\
	&&77.97&E1&1.05(15)$\times$10$^{-5}$ e$^2$fm$^2$&4.3(6)$\times$10$^{-6}$\\
160.73(15)&$<$900 ps&44.6&M1&$>$1.5(5)$\times$10$^{-2}$ $\mu^2_N$&$>$8(3)$\times$10$^{-3}$\\
238.01(15)&57(11)ps&77.17&E1&4.4(12)$\times$10$^{-4}$ e$^2$fm$^2$&1.8(5)$\times$10$^{-4}$\\
	&&121.96&E1$^a$&2.1(5)$\times$10$^{-4}$ e$^2$fm$^2$&9(2)$\times$10$^{-5}$\\
	&&141.88&E1$^a$&$\leq $7.6(19)$\times$10$^{-5}$ e$^2$fm$^2$&$\leq $3.1(8)$\times$10$^{-5}$\\
	&&219.69&E1$^a$&1.2(3)$\times$10$^{-4}$ e$^2$fm$^2$&4.8(12)$\times$10$^{-5}$\\
	&&237.86&M1$^a$&2.5(4)$\times$10$^{-3}$ $\mu^2_N$&1.4(2)$\times$10$^{-3}$\\
266.76(10)&90(20)ps&106.48&M1$^a$&6.4(18)$\times$10$^{-4}$ $\mu^2_N$&3.6(10)$\times$10$^{-4}$\\
	&&150.75&M1$^a$&3.6(10)$\times$10$^{-5}$ $\mu^2_N$&2.0(6)$\times$10$^{-5}$\\
	&&170.41&M1$^a$&4.4(12)$\times$10$^{-5}$ $\mu^2_N$&2.4(7)$\times$10$^{-5}$\\
	&&192.00&E1$^a$&3.7(9)$\times$10$^{-5}$ e$^2$fm$^2$&1.5(4)$\times$10$^{-5}$\\
	&&198.18&E1&1.4(3)$\times$10$^{-4}$ e$^2$fm$^2$&5.6(14)$\times$10$^{-5}$\\
	&&205.00&E1&1.4(4)$\times$10$^{-4}$ e$^2$fm$^2$&5.9(16)$\times$10$^{-5}$\\
	&&228.73&E1$^a$&2.5(6)$\times$10$^{-5}$ e$^2$fm$^2$&1.0(3)$\times$10$^{-5}$\\
	&&260.82&E1$^a$&1.5(4)$\times$10$^{-5}$ e$^2$fm$^2$&6.2(16)$\times$10$^{-6}$\\
471.60(12)&$<$54ps&204.79&M1&$>$1.3(2)$\times$10$^{-2}$ $\mu^2_N$&$>$7.4(13)$\times$10$^{-3}$\\
	&&355.66&M1&$>$2.1(5)$\times$10$^{-4}$ $\mu^2_N$&$>$1.2(3)$\times$10$^{-4}$\\\
	&&396.92&E1$^a$&$>$2.4(9)$\times$10$^{-6}$e$^2$fm$^2$&$>$1.0(4)$\times$10$^{-6}$\\
	&&403.03&E1$^a$&$>$1.2(2)$\times$10$^{-5}$ e$^2$fm$^2$&$>$5.0(10)$\times$10$^{-6}$\\
	&&409.89&E1&$>$4.1(8)$\times$10$^{-5}$ e$^2$fm$^2$&$>$1.7(3)$\times$10$^{-5}$\\
\hline
\hline
\end{tabular}
\par
$^a$ The multipolarity is taken from the decay scheme. For the case of possible E2 or M1 character the latter is chosen if compatible with the level scheme.
\end{table}
\clearpage

\begin{table}[ht]
\caption{Intrinsic dipole moments, $\mid D_0\mid$, for E1 transitions in nuclei near $^{231}$Ac. }
\label{Do-moment}\vspace{0.5cm} \tabcolsep5pt
\begin{tabular}{cccccc}
\hline
\hline
Nucleus & E$_i$ & I$^{\pi}_i$K$_i$ & E$_{\gamma }$ & I$^{\pi}_f$K$_f$ & $\mid D_0\mid $ \\
& (keV) & & (keV) & & (e fm) \\
\hline
$^{227}$Fr & 144.32 & 3/2$^-$3/2 & 141.54 & 3/2$^+$3/2 & 0.034(5)$^a$\\
 & 164.95 & 5/2$^+$3/2 & 162.17 & 3/2$^-$3/2 & 0.089(7)$^a$ \\
\hline
$^{227}$Ac & 27.37 & 3/2$^+$3/2 & 27.37 & 3/2$^-$3/2 & 0.0265(16)$^b$ \\
\hline
$^{231}$Ac & 116.02 & 3/2$^-$3/2 & 54.29 & 3/2$^+$3/2 & 0.0239(11) \\
 &  &  & 47.45 & 5/2$^+$3/2 & 0.0070(6) \\

$^{231}$Ac & 238.01 & 3/2$^+$3/2 & 121.96 & 3/2$^-$3/2 & 0.039(5) \\
 & & & 77.17 & 5/2$^-$3/2 & 0.068(9) \\

$^{231}$Ac & 266.76 & 3/2$^-$3/2 & 205.00 & 3/2$^+$3/2 & 0.032(4)\\
 &  &  & 198.18 & 5/2$^+$3/2 & 0.038(5) \\

\hline
\hline
\end{tabular}

$^a$ D$_0$ moment values taken from~\cite{kur97}.\\
$^b$ B(E1) value taken from~\cite{bro01}.\\

\end{table}
\clearpage

%% Figures
%%

%1

\clearpage

%2

\clearpage

%3

\clearpage

%4

\clearpage

%5

\renewcommand{\thefigure}{\arabic{figure}-\alph{subfigure}}
\setcounter{subfigure}{1}
\begin{figure}
  \begin{center}
\epsfig{file=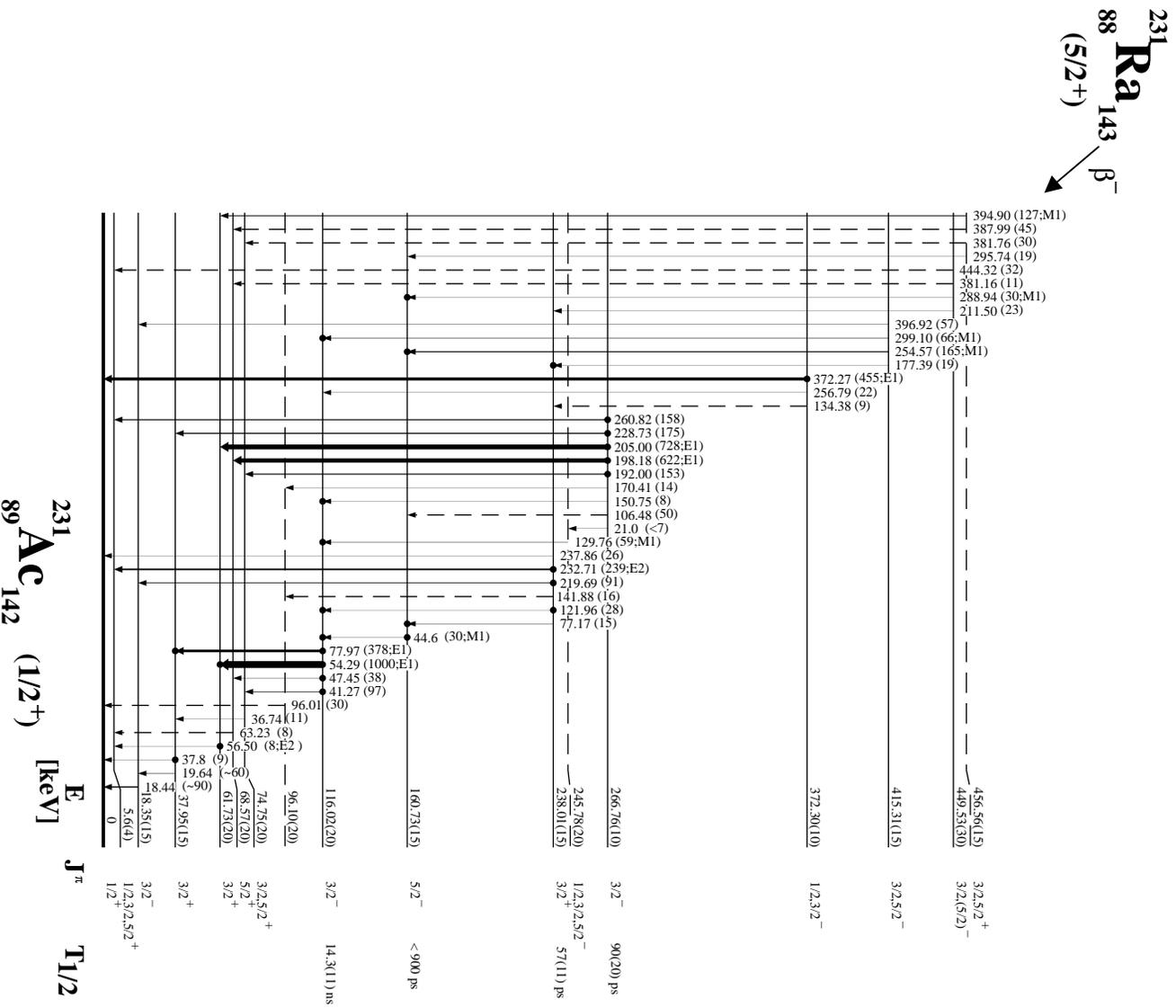,width=14cm}
\caption{Partial level scheme of \nuc{231}Ac. Levels up to 460 keV
excitation energy. Strong coincidences are indicated by dots.}
    \label{esq-1}
  \end{center}
\end{figure}
\clearpage

\addtocounter{figure}{-1}
\addtocounter{subfigure}{1}
\begin{figure}
  \begin{center}
\epsfig{file=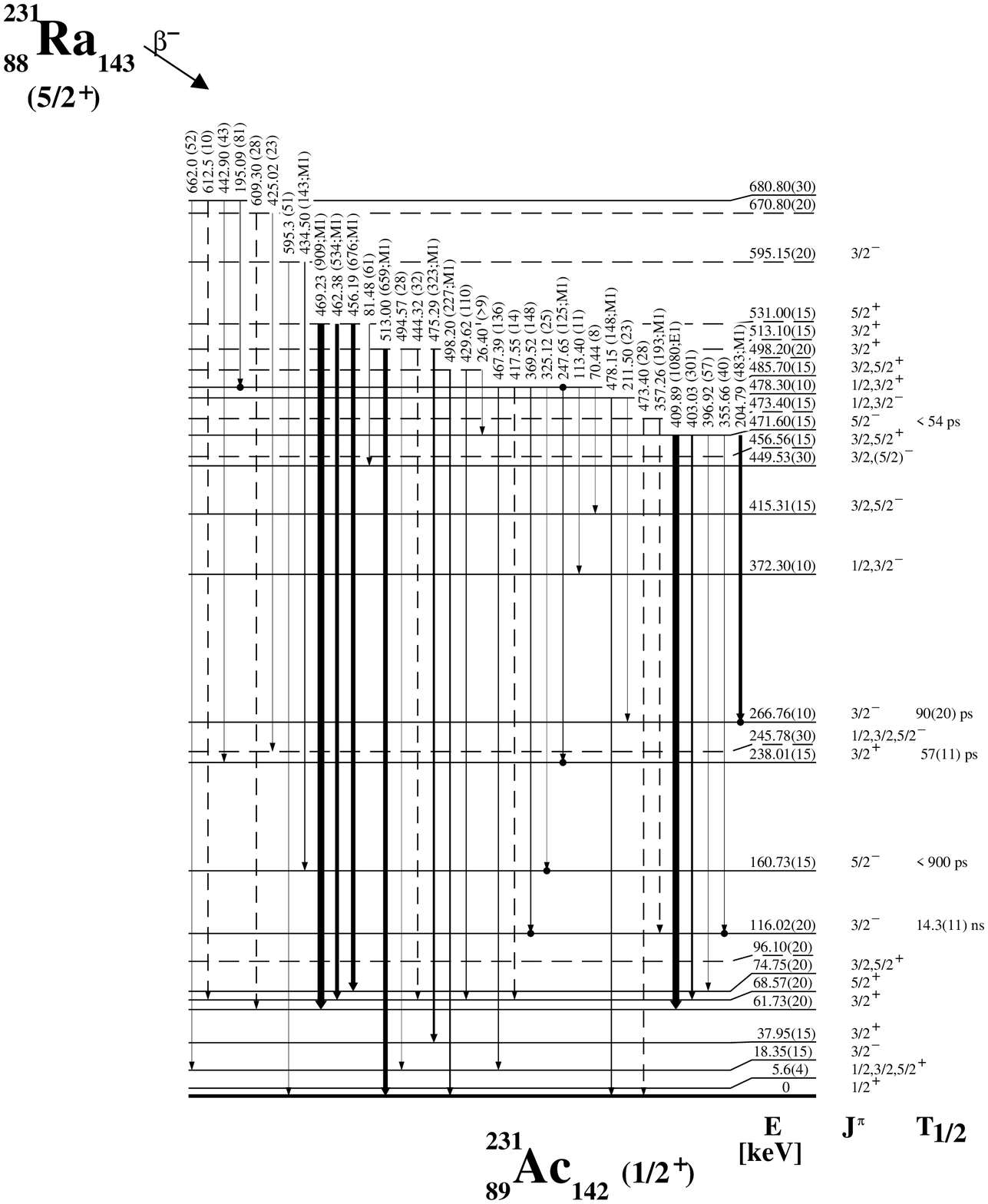, width=14cm}
\caption{Partial level scheme of \nuc{231}Ac.
 Levels up to 681 keV excitation energy
are shown. Strong coincidences are indicated by dots.}
   \label{esq-2}
  \end{center}
\end{figure}
\clearpage

\addtocounter{figure}{-1}
\addtocounter{subfigure}{1}
\begin{figure}
  \begin{center}
\epsfig{file=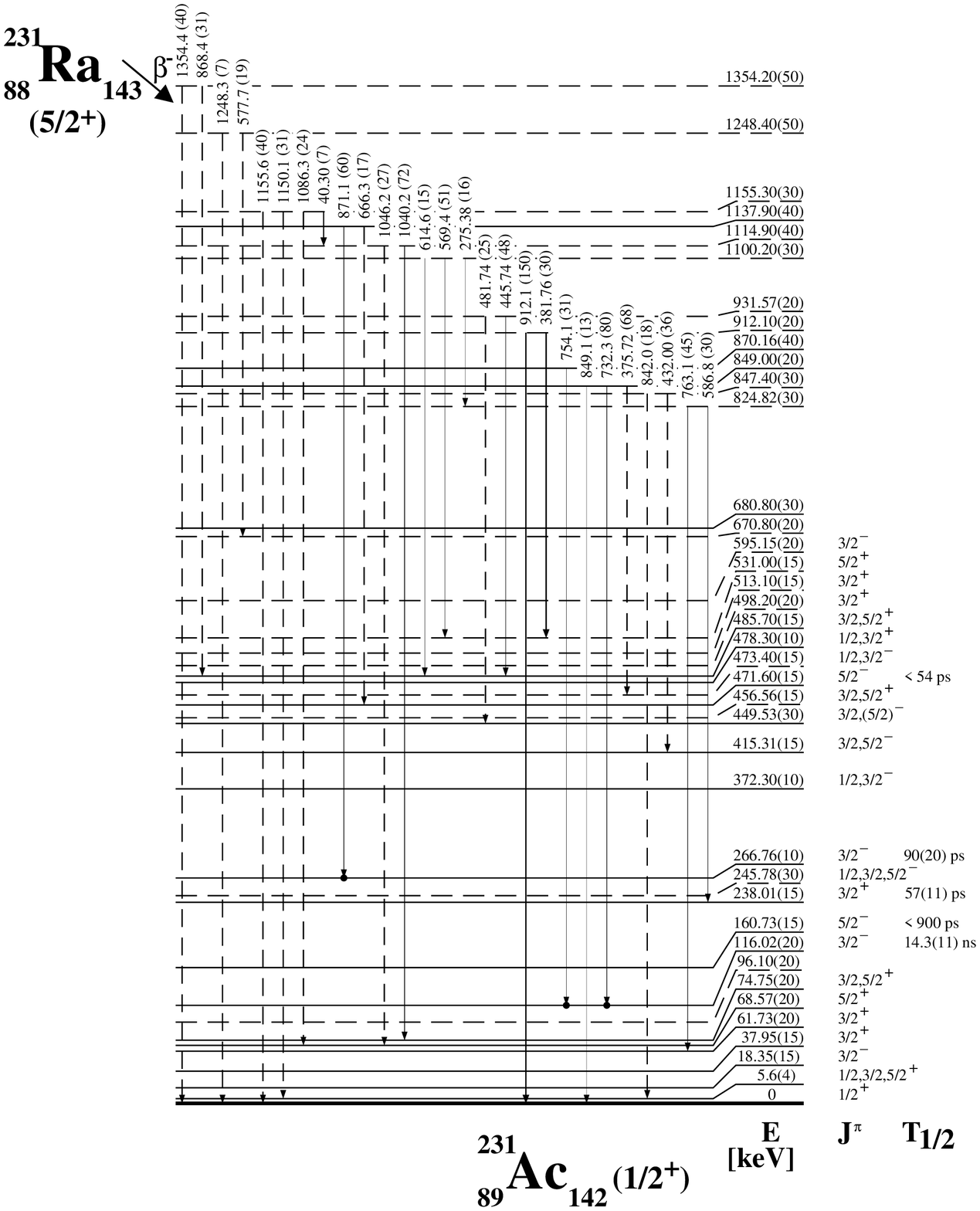, width=14cm}
\caption{Partial level scheme of \nuc{231}Ac.
The upper part of the level scheme is shown. Strong coincidences are indicated by dots.}
   \label{esq-3}
  \end{center}
\end{figure}
\clearpage
\renewcommand{\thefigure}{\arabic{figure}}

%6

 \begin{figure}
  \begin{center}
\epsfig{file=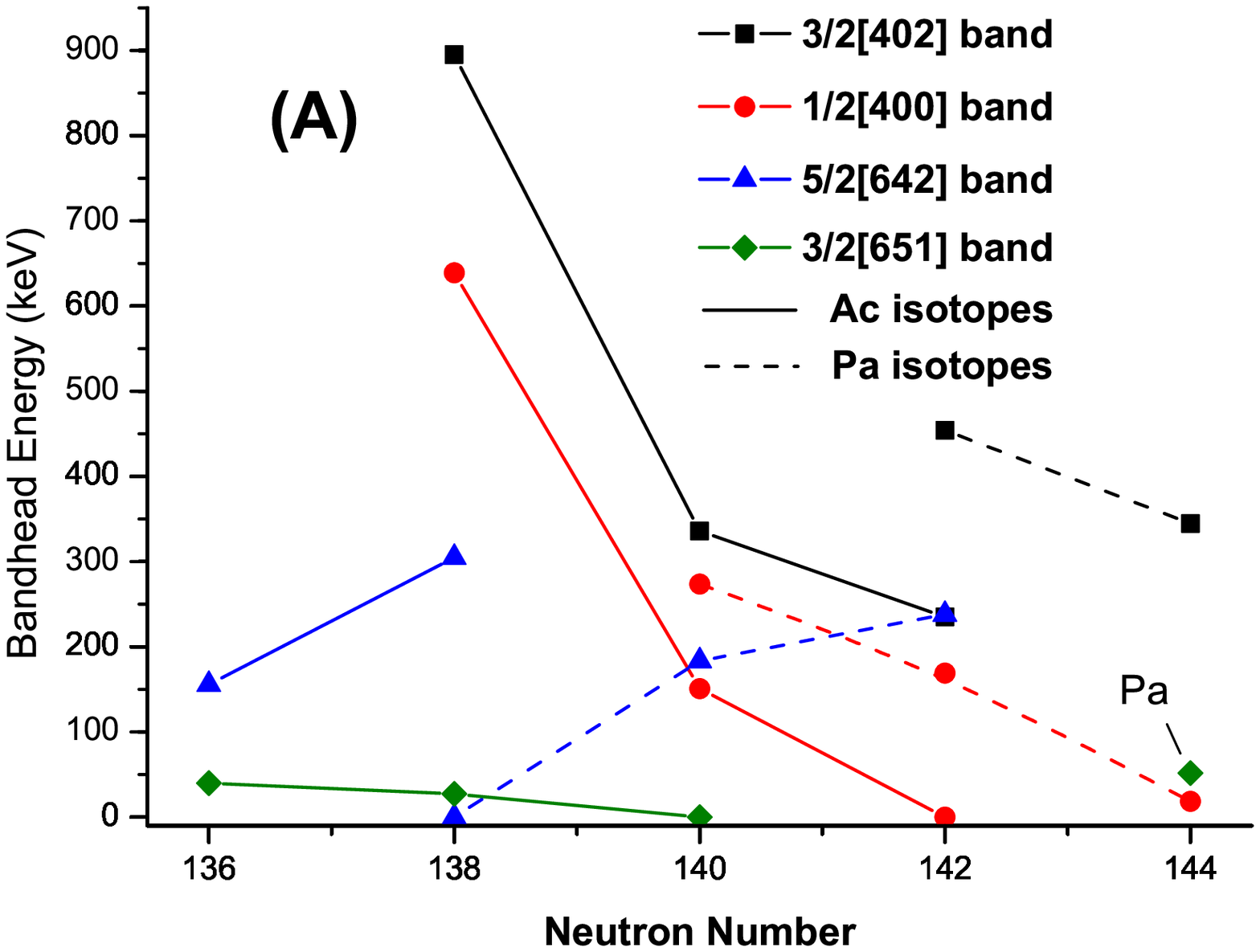, width=11cm}
\epsfig{file=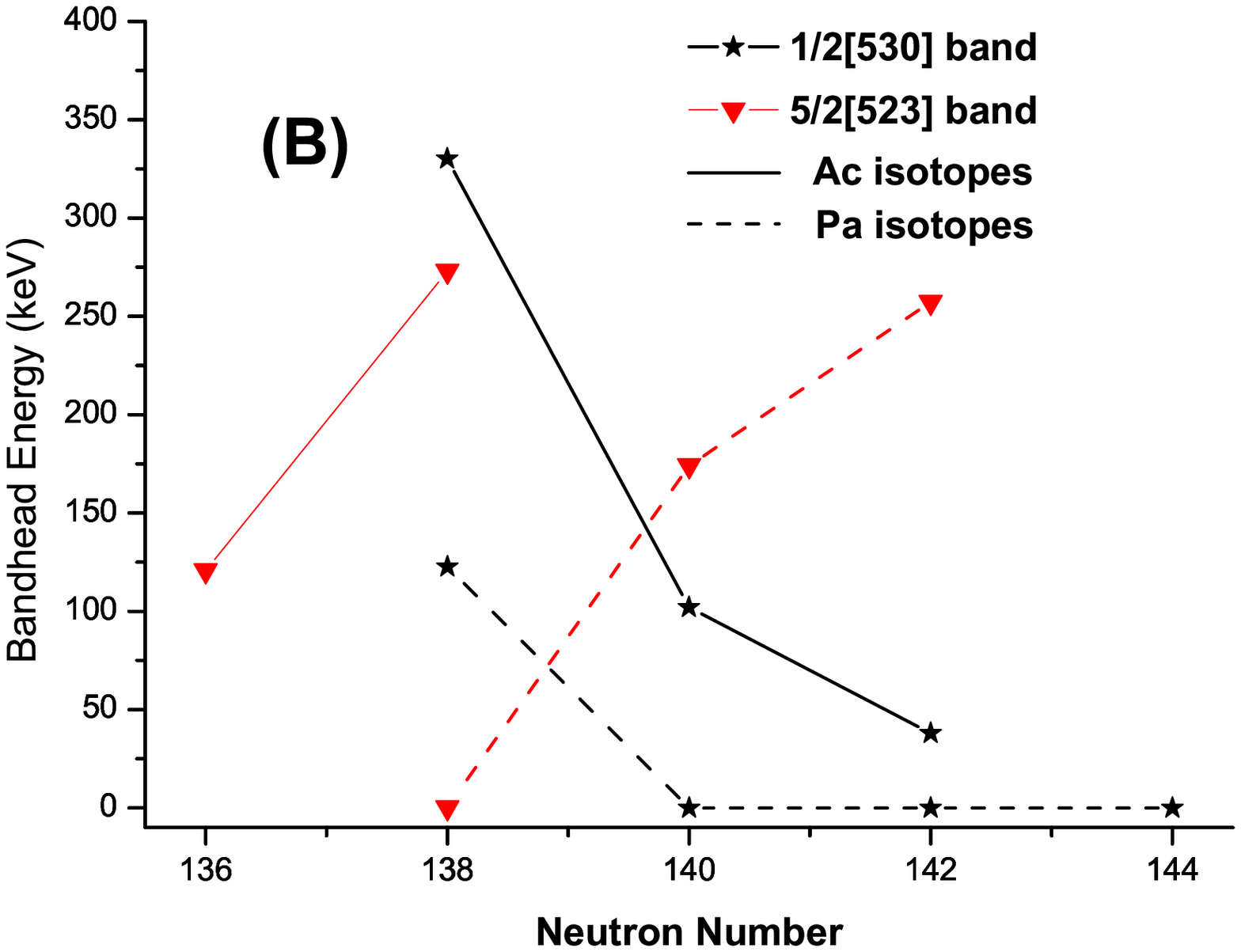, width=11cm}
  \caption{Excitation energy systematics for the 1/2[400], 1/2[530], 3/2[651],
 3/2[402], 5/2[523], 5/2[642] bands in the neutron rich Ac and Pa  isotopes around N=142. The lowest
 state of the 3/2[651] band in $^{231}$Pa is the 5/2$^+$ level at 86.48 keV, 8.18 keV
below the 3/2$^+$. Data are taken from~\cite{bur03,bandas-nucleos}.}
    \label{bandas}
  \end{center}
\end{figure}
\clearpage

%7

\begin{figure}
  \begin{center}
\epsfig{file=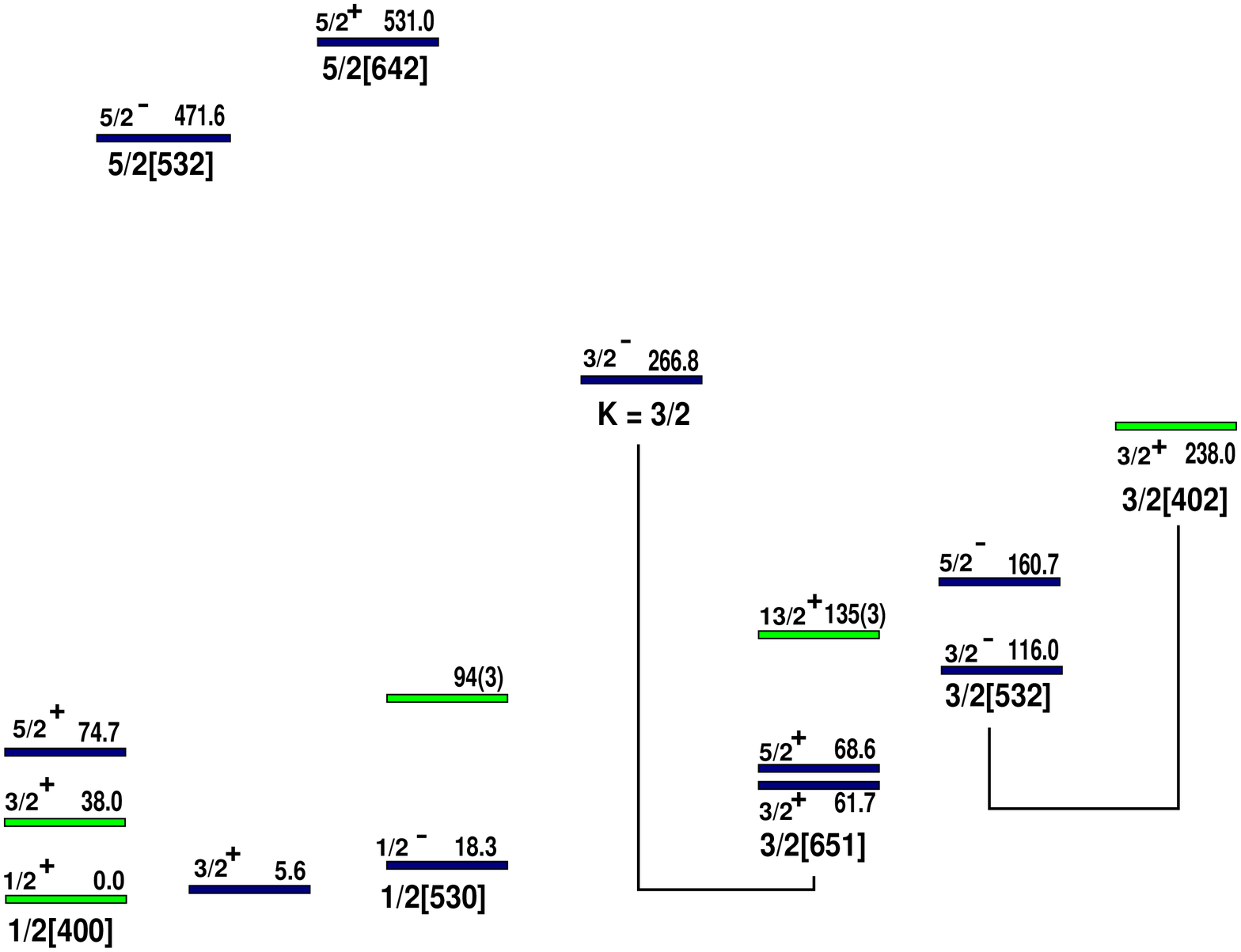, width=14cm}
  \caption{Tentative assignment of the low energy experimental levels
in $^{231}$Ac into rotational bands, see text. The levels in green were originally established in reference \cite{tho77}.}
    \label{bandas-2}
  \end{center}
\end{figure}
%\clearpage

\end{document}